\documentclass[preprint]{aastex}
\usepackage{natbib}
\usepackage{booktabs}
\usepackage{threeparttable}
\usepackage{bbding}
\usepackage{stmaryrd}
\usepackage{longtable}
\usepackage{tabularx}
\usepackage{lineno}

\begin{document}

\title{A Statistical Study of Short-period Decayless Oscillations of Coronal Loops in an Active Region}

\author{Dong~Li\altaffilmark{1} and David~Long\altaffilmark{2}}
\affil{$^1$Purple Mountain Observatory, Chinese Academy of Sciences, Nanjing 210023, PR China \\
       $^2$UCL-Mullard Space Science Laboratory, Holmbury House, Holmbury St. Mary, Dorking, Surrey, RH5 6NT, UK \\
       }
     \altaffiltext{}{Correspondence should be sent to: lidong@pmo.ac.cn}
\begin{abstract}
Coronal loop oscillations are common phenomena in the solar corona,
which are often classified as decaying and decayless oscillations.
Using the high-resolution observation measured by the Extreme
Ultraviolet Imager (EUI) onboard the Solar Orbiter, we statistical
investigate small-scale transverse oscillations with short periods
($<$200~s) of coronal loops in an active region, i.e., NOAA~12965. A
total of 111 coronal loops are identified in EUI~174~{\AA} images,
and they all reveal transverse oscillations without any significant
decaying, regarding as decayless oscillations. Oscillatory periods
are measured from $\sim$11~s to $\sim$185~s, with a median period of
40~s. Thus, they are also termed as short-period oscillations. The
corresponding loop lengths are measured from $\sim$10.5~Mm to
$\sim$30.2~Mm, and a strong dependence of oscillatory periods on
loop lengths is established, indicating that the short-period
oscillations are standing kink-mode waves in nature. Based on the
coronal seismology, kink speeds are
measured to $\sim$330$-$1910~km~s$^{-1}$, and magnetic field
strengths in coronal loops are estimated to about 4.1$-$25.2~G,
while the energy flux carried by decayless kink oscillations lies in
the range from roughly 7~W~m$^{-2}$ to 9220~W~m$^{-2}$. Our
estimations suggest that the wave energy carried by short-period
decayless kink oscillations can not support the coronal heating in
the active region.
\end{abstract}

\keywords{Solar coronal loops --- Solar ultraviolet emission --- Solar
oscillations --- Magnetohydrodynamics --- Solar coronal seismology}

\section{Introduction}
Coronal loops are basic building magnetic structures in solar outer
atmospheres, which are frequently detected in active regions (ARs),
coronal holes, and quiet-Sun regions \citep{Reale14,Mac22}. The
observed lengths of coronal loops are ranging from several hundreds
kilometers (km) to a few hundreds megameters (Mm) in the solar
corona, and the temperature of plasma loops is in the magnitude
order of mega-Kelvin (MK) \citep{Aschwanden11,Peter13,Gupta19}.
Thus, the coronal loop can be easily observed in wavebands of
extreme-ultraviolet (EUV) and Soft X-ray (SXR). Observations find
that a typical coronal loop-like structure generally following the magnetic field
line, for instance, the open coronal structure has only one apparent
polarity and could extend radially along the magnetic field line,
while the closed coronal loop connects to double footpoints that are
rooted in positive and negative polarities, respectively
\citep[e.g.,][]{Watko00,Feng07,Peter12,Li20a}. Therefore, coronal
loops are thought to be associated with the mass supply and plasma
heating in the solar corona, and would help us to understand the
coronal heating problem better \citep[e.g.,][]{Klimchuk15,Li19,Antolin22}.

Coronal loops are highly dynamic. The prominent dynamic is
the transverse oscillation, which was first detected by the
Transition Region and Coronal Explorer \citep[TRACE;][]{Handy99}
spacecraft \citep[see,][]{Aschwanden99,Nakariakov99}. These observed
transverse oscillations are usually decaying rapidly, namely they
can only persist over a few wave periods
\citep{Goddard16,Su18,Nechaeva19}. In the Solar Dynamics Observatory
\citep[SDO;][]{Pesnell12} era, the loop oscillations without
significant damping, regarded as decayless oscillations, are found
to be a common phenomenon in the solar corona
\citep[e.g.,][]{Wang12,Anfinogentov13,Anfinogentov15}. Transverse
decayless oscillations are also detected in coronal bright points
\citep{Gao22}, solar flares \citep{Tian16,Li18a}, and prominence
threads \citep{Diaz01,Ning09,Li18b}, since they all consist of loop-
or thread-like structures that could be regarded as thin magnetic
flux tubes in the corona \citep{Goossens13}. These observations
further confirm that transverse decayless oscillations are
omnipresent in solar atmospheres \citep{Okamoto07,Tian12,Li22a}, and
thus could provide persistent energy input which compensates
energy losses in the solar chromosphere and corona
\citep[e.g.,][]{Morton12,Tian14,Van20}. Decaying oscillations are
often excited by external solar eruptions, such as solar flares,
coronal jets and rains, EUV waves, reconnection outflows, or
flux-rope eruptions, and so on
\citep[e.g.,][]{Antolin11,Zimovets15,Shen17,Shen18a,Reeves20,Zhang20,Zhang22}.
Their apparent displacement amplitudes are much larger than 1~Mm,
while their decaying time is about several oscillatory periods and
decreases with oscillating amplitudes
\citep{Goddard16,Nechaeva19,Zhang20b,Ning22}. Conversely, decayless
oscillations often have small displacement amplitudes, i.e.,
$<$1~Mm, which are always shorter than the minor radius of
oscillating loops \citep{Anfinogentov15}. The driving mechanism of
decayless oscillations is still debated \citep{Mandal22}, such as
the magnetic arcade model \citep{Hindman14}, the Kelvin-Helmholtz
(KH) vortices model \citep{Antolin16}, the self-oscillation model
\citep{Nakariakov16}, the footpoint harmonic-driver model
\citep{Karampelas17}, the steady-flow driver model
\citep{Karampelas20}, and the random-motion driver model
\citep{Afanasyev20,Ruderman21}, etc. Recently, the decayless
transverse oscillations of coronal loops were found to be induced by
a solar flare, which just increases oscillatory amplitudes
\citep{Mandal21}.

Transverse oscillations of coronal loops are often interpreted as
magnetohydrodynamic (MHD) waves in the solar corona, including
decaying and decayless oscillations \citep[see,][and references
therein]{Nakariakov05,Nakariakov20}. In observations of transverse
oscillations, one of the most commonly measured MHD modes is termed
as `kink oscillations', which are always perpendicular to the bulk
parameters of oscillating loops, nonaxisymmetric, and nearly
incompressive in the long-wavelength regime
\citep{Nakariakov21,Nakariakov22,Lopin22}. The standing kink-mode
oscillation may be triggered by an impulsive energy-release process,
i.e., an erupted event in the low corona \citep{Nakariakov21}. The
oscillatory period is found to increase linearly with the length of
plasma loops \citep{Anfinogentov15,Goddard16,Zhong22}. Another type of transverse oscillations in plasma loops is the fast sausage-mode oscillations. Unlike kink waves, they are essentially compressive and strongly dispersive. Wave motions are also perpendicular to the loop axis, but axisymmetric, i.e., expanding-contracting of the loop cross section instead of `kinking' of the loop axis \citep{Chen15,Sadeghi19,Lib20,Li21,Guo21}. For the coronal conditions expected periods of sausage oscillations are about 10 seconds and below. SDO/AIA and previous generations of EUV solar telescopes do not have high enough observational cadence to detect such short-period oscillations. Indeed, imaging observations of sausage oscillations are rare. \cite{Nakariakov03} discussed global sausage oscillations, meaning fundamental-mode waves, when the wavelength is equal to the doubled length of the oscillating loop, and stated that they can only exist in thicker and denser plasma loops. While the higher harmonics and local sausage oscillations can also appear in thin and less dense loops. The
global fast sausage-mode oscillation is seen in a hotter flaring
loop with spectroscopic observations \citep{Tian16}, while the
standing slow and fast sausage waves are simultaneously observed in
a flare loop by using the time series in the Ly$\alpha$ line
emission \citep{Van11}. Slow magnetoacoustic waves, also regarded as
`SUMER' oscillations, are often used to interpret coronal loop
oscillations \citep{Ofman02,Wang11,Wang21}. The slow
magnetoacoustic-mode oscillation is a longitudinal-mode wave, which
is predominantly the parallel flow along plasma loops and always
decaying rapidly and often has a longer oscillatory period
\citep{Kumar13,Yuan15,Kolotkov19,Nakariakov19}. The study of MHD
oscillations in coronal loops becomes a topic of particular
interest, since it is an important tool for remotely inferring
plasma parameters and estimating magnetic field strengths in the
solar corona and chromosphere, named as `MHD coronal seismology'
\citep{Dem12,Yuan16a,Yuan16b,Yuan16c,Long17,Anfinogentov22,Li22a,Pascoe22}.
Moreover, they can allow us to global map the coronal magnetic field
\citep[e.g.,][]{Yang20a,Yang20b}.

Transverse oscillations are usually observed as the spatial
displacement oscillations of plasma loops in EUV and SXR image
sequences
\citep{Aschwanden99,Nakariakov99,Anfinogentov13,Duckenfield18,Li20b,Zhang20b}.
They are also detected as Doppler velocity oscillations of magnetic
loops in EUV spectral lines when the slit happens to cross the
plasma loop \citep{Mariska10,Zhou16,Li17}, which can effectively
avoid the saturation effect of imaging observations. Imaging
observations showed that the oscillatory period in coronal loops was
ranging from a few minutes to several tens minutes
\citep{Anfinogentov15,Shen18b,Duckenfield19,Nechaeva19,Goddard20,Zhang22}.
On the other hand, spectroscopic observations found that the
oscillatory period in hot flaring loops could be less than 1~minute
\citep{Tian16,Li18a}. This is mainly because the time cadence of
EUV/SXR images is a bit lower, for instance, the Atmospheric Imaging
Assembly \citep[AIA;][]{Lemen12} on board SDO takes EUV images at a
cadence of 12~s, making it difficult to detect short-period
oscillations. Now, the high-resolution telescope of the Extreme
Ultraviolet Imager \citep[EUI;][]{Rochus20} onboard the Solar
Orbiter \citep[SolO;][]{Muller20} can provide the higher spatial and
temporal resolution image sequences at the wavelength of 174~{\AA},
giving us an opportunity to investigate the small-scale transverse
oscillations at short periods, i.e., $<$1~minute. In order to
balance the strong energy losses and preserve the high temperature
in the corona, it must be ongoing heated by
$\sim$100$-$200~W~m$^{-2}$ for the quiet-Sun region, while it needs
as high as $\sim$10$^4$~W~m$^{-2}$ for the active regions
\citep{Withbroe77}. In the past few decades, many mechanisms have
been proposed to be possible candidates for the deposition of input
energies, i.e., MHD waves \citep{Van20,Shi21}. In particularly, the
decayless kink-mode wave at short periods could contribute to the
energy input that is required to heat the quiet corona
\citep[cf.][]{Petrova22}. However, it is only the case study, a
statistical study of decayless oscillations at short periods is
still rare, especially for active region loops.

In this work, we present a statistical investigation of decayless
oscillations at short periods and estimate their energy flux using
the MHD coronal seismology. The article is organized as follows:
Section~2 describes in detail observations and instruments used in
this study, Section~3 gives details of the data reduction and MHD
methods, Section~4 presents our statistical results, Section~5
offers discussions, and a brief summary is given in Section 6.

\section{Observations and Instruments}
The observational data sets are mainly from SolO/EUI, which contains
three telescopes: two High Resolution Imagers (HRI) and one Full Sun
Imager (FSI). The labeled `HRI$_{\rm EUV}$' telescope provides local
coronal images with a field of view (FOV) of about
16.8$'$$\times$16.8$'$ at 174~{\AA}, which is dominated by two
coronal lines of Fe IX and Fe X and has a formation temperature of
about 1~MK. The named `HRI$_{\rm Ly\alpha}$' telescope provides
local chromospheric images with the same FOV of
$\sim$16.8$'$$\times$16.8$'$ at 1216~{\AA}, which is characterized
by the Ly$\alpha$ emission of neutral hydrogen line
\citep[cf.][]{Berghmans21,Chen21}. The FSI telescope observes the
full-disk Sun with an uniquely large FOV of about
3.8$^{\circ}$$\times$3.8$^{\circ}$ at wavelengths of 174~{\AA} and
304~{\AA} \citep[cf.][]{Muller20,Mierla22}. The data sets analysed
in this study were obtained on 17 March 2022. On that day, SolO was
located at about 26.5$^{\circ}$ west in solar longitude from the
Sun-Earth line, and the distance between the SolO and the Sun was
about 0.379~AU, as shown in Figure~\ref{location}. Hence, it looked
at the Sun from a different perspective, comparing with the
telescopes in Earth orbit.

The Level~2 data production is downloaded from the EUI Data Release
5.0\footnote{https://doi.org/10.24414/2qfw-tr95}, and they have been
pre-processed by the EUI team (i.e., EUI\_PREP.PRO). We also
performed a cross-correlation technique to test the image
co-alignment \citep[cf.][]{Mandal22}. The targeted active region
(AR) of NOAA12965 was observed by the HRI$_{\rm EUV}$ telescope on
17 March 2022 during 03:18:00$-$04:02:51~UT, as shown in
Figure~\ref{img}~(a). The HRI$_{\rm EUV}$ images at 174~{\AA} were
acquired with a time cadence of about 3~s and a pixel scale of
$\sim$0.49\arcsec, corresponding to about 135~km, as listed in
Table~\ref{tab_instr}. The HRI$_{\rm Ly\alpha}$ images at 1216~{\AA}
were provided from 03:18:00~UT to 03:46:55~UT, the time cadence is
about 5~s, and each pixel scale corresponds to about 282~km. FSI
captured the full-disk images at 304~{\AA} between 03:50:00~UT to
04:03:00~UT with a time cadence of 60~s, and every pixel scale
corresponds to about 1219~km. There is a data gap for the FSI image
acquisition at 174~{\AA}, which missed our observation. It should be
pointed out that the time is selected at the SolO location, which
has a time difference of about 307.5~s with respect to the light
travel time from the Sun to the Earth asset.

The Spectrometer/Telescope for Imaging X-rays
\citep[STIX;][]{Krucker20} on board SolO provides X-ray imaging
spectroscopy of the Sun at a time cadence of 4~s. The energy range
measured by STIX is mainly from 4~keV to 150~keV. Here, we analyzed
the STIX quick-look light curves at energy ranges of 4$-$10~keV,
10$-$15~keV, and 15$-$25~keV, as shown in Table~\ref{tab_instr}. The
AR of NOAA12965 was also observed by SDO/AIA in Earth orbit from a
different perspective, as shown in Figure~\ref{location}. The energy
spectrum and X-ray images are not used due to the lower X-ray
counts. AIA is designed to measure full-disk solar images at a
spatial scale of 0.6\arcsec\ per pixel (corresponds to about
$\sim$435~km) in seven EUV and two UV wavelength bands. It also
provides rebinned solar images at a lower spatial scale and a longer
time cadence \citep{Lemen12}. In our study, two AIA images at
wavelengths of 171~{\AA} and 131~{\AA} are used, and they are
dominated by coronal and flare lines, such as Fe IX ($\sim$0.6~MK),
Fe VIII ($\sim$0.4~MK) and Fe XXI ($\sim$11~MK), respectively. The
AIA 171~image has a time cadence of 12~s and a pixel scale of about
435~km, while the 131~{\AA} image has been spatially rebinned
(4$\times$) with a time cadence of 120~s and a pixel scale of about
1740~km, as shown in Table~\ref{tab_instr}.

\section{Data reduction and Methods}
We analyzed fine-scale coronal loops (e.g., the loop length is a few
tens Mm) in the AR of NOAA~12965, as shown in Figure~\ref{img}.
Panel~(a) presents the targeted AR at 03:18:00~UT in HRI$_{\rm
EUV}$~174~{\AA} observed by SolO/EUI, and many groups of loop-like
structures appear in the AR, including large-scale (e.g., the length
order of about one hundred Mm) and fine-scale coronal loops. Those
large-scale loops can also be seen in the AIA~171~image that
observed the targeted AR at a different perspective, as shown in
panel~(b). Here, we have corrected the time shift between SDO/AIA
and SolO/EUI, i.e., 307.5~s. On the other hand, these fine-scale
loops can be clearly seen in the HRI$_{\rm EUV}$~174~{\AA} image, as
indicated by R1 and R2 regions. However, they are not so obvious in
the AIA~171~image, largely due to its lower spatial resolution. In
Figure~\ref{img}~(c) and (d), we display two sub-fields
corresponding to the green boxes outlined in panel~(a), which
clearly reveal these fine-scale coronal loops. Then, we extracted
thirteen artificial slits that are almost perpendicular to coronal
loops to perform the time-distance map, as indicated by the cyan
lines.

\subsection{Detection of short-period decayless oscillations}
From the online animation, we can see that coronal loops exhibit
highly dynamic, such as drifting motions, rotation movements,
transverse and longitudinal oscillations. Here, we focus on the
transverse oscillations of fine-scale coronal loops, particularly
for the decayless oscillations with periods shorter than 200~s. Some
brightest loops are not selected to avoid the saturated pixels.

In order to enhance the appearance of small-scale oscillations, a
smooth window of 200~s is applied to the original map
\citep[cf.][]{Mandal22,Zhong22}. The smooth window is reasonable,
since we are only interested in the short periods, i.e., $<$200~s.
Figure~\ref{slt1}~(a) plots the time-distance map after removing a
smooth version along the slit S1 in Figure~\ref{img}~(c), which are
made from the image sequences at HRI$_{\rm EUV}$~174~{\AA}. Here,
the starting point of the time-distance analysis is indicated by the
green symbol of `$\ast$'. Totally, 13 coronal loop oscillations (the
detail can be seen in the Appendix~A) are found in this
time-distance map, as indicated by cyan circles, which represent the
central or boundary positions of oscillating loops. Generally, the
loop centers are determined by Gaussian fitting
\citep[e.g.,][]{Wang12,Zhong22}. However, it could be difficult to
apply this method when a series of overlapping loop structures
simultaneously appear in the time-distance map
\citep{Anfinogentov15,Goddard16,Gao22}. Thus, we manually identified
those brightest/weakest pixels as loop centers or edges. It should
be pointed out that although some coronal loops reveal oscillations,
they also show strongly vortex motions, and we have to abandon them,
i.e., those coronal loops in the region centered at x$\approx$1500~s
and y$\approx$15~Mm. These thirteen coronal loops show transverse
oscillations without significant decaying, which could be regarded
as decayless oscillations. We also note that some coronal loop
oscillations drift significantly, such as loops 6 and 13, while some
others drift weakly, i.e., loops 4 and 7. Therefore, the decayless
oscillation is fitted by a widely adopted function, namely, the sine
function with a linear trend but without the decaying term
\citep[cf.][]{Anfinogentov13,Anfinogentov15,Li20b,Zhang20b,Gao22},
as indicated by~Equation~\ref{eq1}:

\begin{equation}
  A(t)=A_m \sin(\frac{2 \pi}{P}~t+ \phi )+kt+A_0,
\label{eq1}
\end{equation}
where $A_m$ stands for the displacement amplitude, $P$ is the
oscillatory period, $\phi$ and $A_0$ represent the initial phase and
initial position of the oscillating motion, while $k$ is constant
and denotes to the drifting velocity of the oscillating loop in the
plane-of-the-sky \citep[e.g.,][]{Ning09,Li18b,Zhang22}. The fitting
results are indicated by the magenta curves, and they appear to
match well with the oscillatory positions of coronal loops. Using
the derivative of Equation~\ref{eq1}, the velocity amplitude ($v_m$)
of the decayless oscillations could be derived
\citep[cf.][]{Li22a,Petrova22}, that is, $v_m=2\pi A_m/P$. Some key
oscillatory parameters of loop oscillations are listed in
Table~\ref{tab_loop1}.

In order to take a closer look at the oscillating loops in the
corona, we arbitrarily extracted the HRI$_{\rm EUV}$~174~{\AA}
images to show fine-scale coronal loops during the time interval of
decayless oscillations, as indicated by two red vertical lines in
Figure~\ref{slt1}~(a), which happens to cross the oscillating
loops~1, 6 and 12. In Figure~\ref{slt1}~(b) and (c), we show EUV
images at HRI$_{\rm EUV}$~174~{\AA} at two fixed instances of time,
which contain three oscillating loops, as marked by red curves. The
cyan line represents the slit center, which has a constant width of
about 675~km (5 pixels), and it is almost perpendicular to the
oscillating loop. The green symbol of `$\ast$' indicates the
starting point of the time-distance analysis in panel~(a). Based on
the HRI$_{\rm EUV}$~174~{\AA} images displayed in
Figure~\ref{slt1}~(b) and (c), the distance between double
footpoints of the coronal loop can be estimated. If we assume a
semi-circular profile for the coronal loop
\citep[cf.][]{Tian16,Gao22,Li22b,Mandal22} and consider the
projection effect owing to the derived position
\citep[e.g.,][]{Aschwanden02,Kumar13,Li17b}, the deprojected loop
length ($L$) could be estimated, as listed in Table~\ref{tab_loop1}.

Using the same method (see details in the Appendix~A), we manually
extracted 6 coronal loop oscillations from slit S2 in region~R1 and
10 coronal loop oscillations from slit S3 in region~R2, as shown in
Figures~\ref{slt2} and \ref{slt3}, respectively. The oscillatory
positions of these coronal loop oscillations are manually marked by
cyan circles, and they are well fitted by Equation~\ref{eq1}, as
indicated by magenta curves, implying that they are all decayless
oscillations. Some oscillating loops are also shown in the HRI$_{\rm
EUV}$~174~{\AA} images (see panels (b) and (c) in Figures~\ref{slt2}
and \ref{slt3}), which have been outlined by red curves. Then, the
loop length can also be estimated by assuming the semi-circular
shape \citep[cf.][]{Tian16,Gao22,Mandal22}. Coronal loop
oscillations with periods longer than 200~s are not considered in
this study, for instance, the coronal loop oscillation in the region
at around x$\approx$500$-$1000~s and y$\approx$10~Mm in
Figure~\ref{slt3}~(a).

At last, a total of 111 coronal loop oscillations are extracted from
thirteen slits in the two regions R1 and R2 using the HRI$_{\rm
EUV}$~174~{\AA} observation. Table~\ref{tab_loop1} lists some key
oscillatory parameters based on the fitting results, i.e., the loop
length, the oscillatory period, the displacement amplitude, and the
velocity amplitude.

\subsection{MHD coronal seismology}
Assuming that all those decayless oscillations of coronal loops are
fundamental mode of standing kink waves, the kink speed ($c_{k}$) of
transverse oscillations could be determined by the loop length
($L$) and the oscillatory period ($P$), as shown in
Equation~\ref{eq2}:

\begin{equation}
  c_{k}~\approx~\frac{2L}{P},
\label{eq2}
\end{equation}

Then, the magnetic field strength ($B$) inside oscillating loops
could be roughly expressed in Equation~\ref{eq3}
\citep[e.g.,][]{Nakariakov03,Nistico13,Li17,Anfinogentov19,Gao22,Tan22,Zhang22}:

\begin{equation}
  B~\approx~c_{k}~\sqrt{\frac{1+\rho_e/\rho_i}{2}}~\sqrt{\mu_0\widetilde{\mu}\rho_i}~,
\label{eq3}
\end{equation}
Where $\mu_0$ represents the magnetic permittivity of free space,
$\mu_0=4\pi~\times~10^{-7}$~N~A$^{-2}$ in the international system
of units (SI), and $\widetilde{\mu}~\approx~1.27$ stands for an
average molecular weight in the solar corona
\citep[e.g.,][]{Nakariakov01,Zhang20}. $\rho_e$ and $\rho_i$ are the
external and internal plasma densities of the coronal loop,
respectively. In our study, a typical value of the internal plasma
density of the coronal loop is applied, such as
$\rho_i~=~1.67~\times~10^{-12}$ kg m$^{-3}$
($\sim$10$^{9}$~cm$^{-3}$), and the density contrast
($\rho_e/\rho_i$) is assumed to be equal to 1/3
\citep[cf.][]{Gao22,Petrova22}.

Next, we can estimate the wave energy flux of the observed coronal
loop oscillations based on the MHD wave seismology \citep[see,][for
the review articles]{Nakariakov05,Nakariakov20}. Using
Equations~\ref{eq4} and \ref{eq5}, the time-averaged wave energy
flux ($E$) carried by the kink-mode MHD wave could be calculated in
the oscillating loop
\citep[e.g.,][]{Goossens13,Van14,Yuan16a,Li22a,Petrova22}:

\begin{equation} 
  b~\approx~A_m\frac{\pi}{L}B,
\label{eq4}
\end{equation}
\begin{equation}
  E~\approx~\frac{1}{4}c_{k}~(\rho_i~v_m^{2}+\frac{b^2}{\mu_0}),
\label{eq5}
\end{equation}
where $b$ stands for the Lagrangian perturbation of magnetic fields,
namely magnetic field perturbation, which could be estimated by the
Lagrangian displacement vector \citep[cf.][]{Yuan16a}, $A_m$ and
$v_m$ represent the displacement and velocity amplitudes of
decayless oscillations, and $L$, $B$, $c_{k}$, and $\rho_i$ are the
loop length, the magnetic field strength, the kink speed, and the
plasma density of oscillating loops, respectively. The estimated
parameters of decayless oscillations are listed in
Table~\ref{tab_loop1}. {A detailed estimation for the energy
flux ($E$) is given in Appendix~B.

\section{Statistical results}
The observed oscillations in fine-scale coronal loops do not show
any systematic decaying, which could be regarded as decayless
oscillations. Their main oscillatory parameters are measured, and
the key physical parameters are also estimated based on the MHD
coronal seismology, as listed in Table~\ref{tab_loop1}. In order to
display the distribution of these parameters, we then draw their
histograms, as shown in Figures~\ref{sta1} and \ref{sta2}. Next, we
will analyze them in details.

\subsection{Distribution of key parameters}
In total, 111 decayless oscillations are analyzed here, which
allowed us to establish the distribution of some key parameters.
Figure~\ref{sta1} presents histograms of the loop length (a), the
oscillatory period (b), the displacement amplitude (c), and the
velocity amplitude (d). Lengths of oscillating loops are measured in
the range between about 10.5~Mm and 30.2~Mm, with a median length of
14.4~Mm and an averaged length of 15.1~Mm. They are much shorter
than previous measurements of several hundreds Mm in coronal loop
oscillations
\citep[e.g.,][]{Aschwanden02,Anfinogentov13,Anfinogentov15,Goddard16,Nechaeva19},
while our estimations are quite similar to those measured in coronal
bright points with a length range from 14~Mm to 42~Mm
\citep[cf.][]{Gao22}. Oscillatory periods are in the range of
$\sim$11$-$185~s, with a median/averaged period of 40/49~s, which is
similar to that detected in decayless oscillations of shorter
coronal loops \citep{Petrova22}, but they are smaller than those in
longer coronal loops
\citep{Nakariakov99,Anfinogentov15,Goddard16,Mandal22}. Displacement
amplitudes are found to range from roughly 51~km to 488~km, with a
median/averaged value of 156/186~km, which is roughly equal to that
(170~km on average) in decayless oscillations of coronal loops, but
it is larger than those of decayless oscillations in coronal bright
points, i.e., ranging from about 27~km to 133~km with an average of
65~km \citep{Gao22}. Velocity amplitudes are measured in about
5.3$-$118.8~km~s$^{-1}$, with a median/averaged speed of
25.6/29.8~km~s$^{-1}$. Our measurements are similar to those
observed in high-frequency decayless waves in coronal loops
\citep{Petrova22}. However, they are much larger than previous
findings (several km~s$^{-1}$) obtained in decayless oscillations of
loop-like structures
\citep[e.g.,][]{Tian12,Anfinogentov13,Nakariakov16,Gao22,Li22a}.

Figure~\ref{sta2} shows histograms of the kink speed (a), the
magnetic field strength (b), the magnetic field perturbation (c),
and the energy flux (d). Kink speeds of oscillations loops are
$\sim$330$-$1910~km~s$^{-1}$ with an averaged speed of
780~km~s$^{-1}$. Magnetic field strengths in oscillating loops are
estimated to about 4.1$-$25.2~G with an averaged strength of 9.8~G,
which is similar to previous findings using the MHD coronal
seismology in coronal loops, i.e., a few Gs to several tens Gs
\citep[e.g.,][]{Nakariakov01,Aschwanden02,Yang20b,Gao22,Zhang22}.
Magnetic field perturbations in coronal loops are found to be about
0.07$-$1.5~G with an averaged perturbation of 0.37~G. Finally,
energy fluxes taken by decayless oscillations are estimated to range
from roughly 7~W~m$^{-2}$ to 9220~W~m$^{-2}$ with an averaged flux
of 820~W~m$^{-2}$, the median flux is only 380~W~m$^{-2}$, implying
the lower energy is dominant. The oscillatory and physical
parameters estimated from decayless oscillations are summarized in
Table~\ref{tab_loop2}.

Figure~\ref{sta2}~(d) shows that the energy flux carried by
decayless oscillations has a broad range, and they concentrated on
the lower energy-end. Hence, we plot the frequency distribution of
energy fluxes in log-log space, as shown in Figure~\ref{sta3}. It
can be seen that the frequency distribution of energy fluxes appears
to follow a power-law behavior, which could be expressed by
Equation~\ref{eq6}:

\begin{equation}
  \frac{dN}{dE}~=~{\rm C}E^{-\alpha},
\label{eq6}
\end{equation}
where $E$ represents the energy flux carried by decayless
oscillations, dN is the number of events in the energy interval [E,
E+dE], C and $\alpha$ (power-law index) stands for two constants,
which are determined by the maximum likelihood estimation
\citep[e.g.,][]{Clauset09,Verbeeck19,Lu21}. This maximum likelihood
method automatically returns a break of the power-low model, as
indicated by the vertical line in Figure~\ref{sta3}. The dropping of
the frequency distribution at higher frequencies is mainly due to
the observational fact that a large number of decayless oscillations
carried weaker energies are ignored in this study, because we only
study these decayless oscillations with periods shorter than 200~s.
Therefore, the power-law index is as hard as about 1.12, which is
far away from the theoretical expectation, i.e., slightly less than
2.0 \citep[e.g.,][]{Hudson91,Crosby93,Su06,Ning07,Li13,Li16,Ryan16}.

\subsection{Statistical scaling between key parameters}
In this subsection, we will establish a statistically significant
relationship between several key parameters, as shown in
Figure~\ref{relat}. The displacement and velocity
amplitudes showed weakly dependence on oscillatory periods or loop
lengths \citep[e.g.,][]{Nakariakov16,Gao22}, and thus we do not
analyze their statistical scaling in this study.

Figure~\ref{relat}~(a) presents the scatter plot between oscillatory
periods and loop lengths in norm-norm space. Similar to previous
findings \citep{Anfinogentov15,Goddard16}, the oscillatory period
linearly (indicated by a magenta line) increases with the length of
oscillating loops. Their linear Pearson correlation coefficient
(cc.) is as high as 0.98, confirming their strong correlation.
Figure~\ref{relat}~(b) shows the scatter plot between oscillatory
periods and energy fluxes in log-log space. We find that, despite
significant scattering of the two parameters, the energy flux
decreases with the growth of the oscillatory period. The negative
value of the Pearson correlation coefficient (-0.77) indicates their
negative correlation, and the magenta line represents a linear fit.

In Figure~\ref{relat}~(c) we show scatter plots between velocity
(black diamonds) and displacement (cyan diamonds) amplitudes and
energy fluxes in log-log space. It can be seen that the energy flux
linearly (indicated by a magenta line) increases with the speed of
velocity amplitudes. However, it is almost not dependent on the
displacement amplitude, which shows quite large scattering. The
Pearson correlation coefficients confirm that the energy flux is
strongly dependent (i.e., 0.97) on the velocity amplitude, but it
has weakly dependence (i.e., 0.23) on the displacement amplitude.
Figure~\ref{relat}~(d) presents scatter plots between magnetic field
perturbations (black squares) or strengths (cyan squares) and energy
fluxes in log-log space. From which, we can see that the energy flux
linearly increases with the magnetic field perturbation, as
indicated by the magenta line, and their Pearson correlation
coefficient is as high as 0.97. Although some scattering, the energy
flux appears to increase with the magnetic field strength, and the
Pearson correlation coefficient between them is 0.75.

\section{Discussions}
Thanks to the high-temporal resolution (i.e., $\sim$3~s) of SoLO/EUI
at the passband of HRI$_{\rm EUV}$~174~{\AA}, 111 short-period
decayless oscillations are observed, which have a median period of
40~s. The short-period decayless oscillations have not been detected
by SDO/AIA, largely due to their low-temporal resolution, such as
12~s. Due to the small-scale amplitude of decayless oscillations,
the motion magnification algorithm is often applied to  increase the
oscillatory amplitude when using the SDO/AIA data
\citep[e.g.,][]{Anfinogentov16,Li20b,Gao22}. On the other hand, the
small-scale decayless oscillations could be directly detected by
SoLO/EUI at the passband of HRI$_{\rm EUV}$~174~{\AA}, because its
high spatial resolution, i.e., 135~km~pixel$^{-1}$ in this study.
The decayless oscillations are simultaneously detected by SDO/AIA at
171~{\AA} and SoLO/EUI at HRI$_{\rm EUV}$~174~{\AA}, suggesting that
the HRI$_{\rm EUV}$~174~{\AA} images could be used to study the
decayless oscillations \citep{Mandal22,Zhong22}, especially for the
decayless oscillations at short periods, i.e., $<$60~s
\citep{Petrova22}. However, the short-period decayless oscillations
correspond to shorter oscillating loops, and these shorter
oscillating loops often reveal highly dynamic, such as fast
drifting motions, as shown in the animation of eui\_174.mp4.
Therefore, they could be observed for only persisting over a few
wave periods, that is, they can not be seen when the oscillating
loop drift away from the slit. In a word, the short-period decayless
oscillations are credible \citep{Petrova22,Zhong22}. Next, we will
discuss them in details.

\subsection{MHD mode of detected decayless oscillations}
Transverse oscillations of coronal loops were first discovered in
1999 by TRACE \citep{Aschwanden99,Nakariakov99}, and they raised a
great interest in connection with MHD waves and the coronal
seismology
\citep[e.g.,][]{Lib20,Nakariakov20,Van20,Yang20a,Yang20b,Nakariakov21,Pascoe22,Zhong22}.
Transverse loop oscillations have been classified into decaying and
decayless oscillations \citep[e.g.,][]{Nistico13}. In our
observations, the decayless oscillations could also be explained by
MHD waves. The slow magnetoacoustic wave can be first excluded
because it belongs to the longitudinal-mode wave and is often
parallel to the oscillating loop
\citep{Wang11,Yuan15,Kolotkov19,Nakariakov19}. Moreover, the local
sound speed ($v_s\approx152\sqrt{\frac{T}{MK}}$) in the coronal loop
\citep[cf.][]{Nakariakov01,Kumar13,Li17} is estimated to be about
152~km~s$^{-1}$, considering that our observations are taken from
the HRI$_{\rm EUV}$~174~{\AA} channel that has a formation
temperature of about 1~MK. The local sound speed is much slower than
the estimated kink speed, as can be seen in
Table~\ref{tab_loop2}.

Kink speeds are estimated to be 330$-$1910~km~s$^{-1}$, with an
averaged speed of 780~km~s$^{-1}$. The median speed is about
710~km~s$^{-1}$, which is roughly equal to the averaged speed. The
estimated kink speeds are much smaller than that requires for the global
sausage-mode wave. For this kind of wave, the speed is expected to
be in the range of $\sim$3000$-$5000~km~s$^{-1}$
\citep[e.g.][]{Nakariakov03,Inglis08}. Moreover, the global sausage
wave is generally seen in the thicker and denser plasma loop,
namely, the density contrast inside and outside the plasma loop
should be large enough, i.e., a density contrast of 42 in the
flaring loop \citep[cf.][]{Tian16}. Obviously, the oscillating loops
in our study are not satisfied with such high density contrast, although
we cannot determine them due to the limited observation. But
previous observations have suggested that the typical density
contrast of coronal loops are about 2$-$10
\citep{Aschwanden02,Gao22,Petrova22}.

Herein, the measured speeds are very close to the typical kink
speed of global kink-mode waves \citep{Nakariakov21}. The estimated
magnetic field strengths are in the range of 4.1$-$25.2~G, with an
average of 9.8~G. Our estimations based on the MHD coronal
seismology are quite similar to previous findings derived from the
kink oscillation of coronal loops, i.e., several Gs to a few tens Gs
\citep{Nakariakov01,Aschwanden02,Long17,Zhang22}. Therefore, the
small-scale transverse oscillations observed in coronal loops could
be regarded as decayless kink-mode oscillations. Moreover, a strong
correlation is found between oscillatory periods and loops lengths
(Figure~\ref{relat}~(a)), confirming the explanation of decayless
oscillations as the standing kink-mode wave in coronal loops.

\subsection{Contribution to the coronal heating}
Benefitting from the high-temporal resolution (i.e., $\sim$3~s)
imaging observations observed by SolO/EUI at the HRI$_{\rm
EUV}$~174~{\AA} passband, the averaged period of decayless kink
oscillations is estimated to 49~s, while the median period is
only 40~s. It is shorter than previous detections in kink
oscillations of coronal loops observed by SDO in AIA~171~{\AA},
namely, hundreds of seconds \citep[see,][for statistical
reults]{Anfinogentov15,Goddard16,Nechaeva19}. Thus, the decayless
kink oscillations are regarded as short-period oscillations
\citep[cf.][]{Petrova22}. Based on the observational fact that
oscillatory periods of kink oscillations are always dependent on
their loop lengths \citep{Anfinogentov15,Goddard16,Nechaeva19}, the
short-period kink oscillations should correspond to shorter loops,
which is consistent with our measurements (i.e., the
averaged loop length is estimated to 15.1~Mm and the median loop
length is 14.4~Mm). The measured loop lengths are much shorter than
previous findings in coronal loop oscillations with longer periods,
which could be as large as several hundreds Mm
\citep[e.g.,][]{Aschwanden02,Anfinogentov13}.

The averaged displacement amplitude of decayless oscillations is
186~km, with a median value of 156~km. This is quite close to
previous findings (170~km on average) in decayless kink oscillations
\citep{Anfinogentov13,Anfinogentov15}. Obviously, the apparent
displacement amplitudes do not gradually decrease with the shorten
of oscillatory periods, which agrees with previous findings, for
instance, the oscillating amplitude does not reveal any dependence
on only one parameter such as oscillatory periods or loop lengths,
but it could be determined by both of them
\citep{Nakariakov16,Gao22,Petrova22}. The displacement amplitudes
measured here are mostly equal to the pixel size of HRI$_{\rm
EUV}$~174~{\AA} (135~km) images, which are reasonable. This is
because sub-pixel displacement amplitudes have been reported in
decayless oscillations in AIA~171~images, and they are demonstrated
to be reliable
\citep[see,][]{Anfinogentov16,Anfinogentov22,Zhong21,Gao22}. The
short oscillatory period would lead to a fast velocity amplitude. As
being expected, velocity amplitudes of decayless oscillations are
found to 5.3$-$118.8~km~s$^{-1}$ with an averaged speed of
29.8~km~s$^{-1}$, which are similar to that of high-frequency
decayless waves \citep{Petrova22}. On the other hand, the velocity
amplitude in our study is much larger than previous observations of
decayless oscillations observed by SDO/AIA, and it seems to exceed
the velocity amplitude of decaying oscillations. Therefore, the
previous classification \citep{Nistico13} such as low-amplitude
decayless and high-amplitude decaying based on velocity amplitudes
is not applicable in this work, and we regarded them as small-scale
decayless oscillations.

Kink-mode MHD waves, one of the possible candidates for coronal
heating, have been intensively studied since their first detection
\citep[see,][for reviews]{Van20,Nakariakov21}. Decayless kink
oscillations are found to be persistent and omnipresent in solar
atmospheres, i.e., coronal loops
\citep{Anfinogentov13,Anfinogentov15}, prominence threads
\citep{Okamoto07,Li22a}, coronal bright points with loop-like
profiles \citep{Gao22}. Thus, they could provide continuous energy
input that is required to heat the solar corona. However, previous
observations found that the dissipating energy carried by decayless
kink oscillations were not enough to heat the ambient plasmas in the
solar corona \citep{Klimchuk15,Li22a,Gao22}. On the other hand, a
recent high-temporal observation from HRI$_{\rm EUV}$~174~{\AA}
images indicates that the short-period (or high-frequency) decayless
kink oscillations take significant energy to supply for heating
the quiet corona, even if they only consider the kinetic energy
\citep[cf.][]{Petrova22}.

Given simultaneously consideration of the kinetic and magnetic
energies (see, Equation~\ref{eq5}), we estimate the energy flux
carried by decayless kink oscillations at shorter periods in an
active region using the HRI$_{\rm EUV}$~174~{\AA} images with a time
cadence of 3~s. It should be pointed out that we have
assumed that the decayless kink oscillations had the same damping
mechanism with decaying kink oscillations. And thus, the estimated
energy fluxes could also regard as energy losses from decayless kink
oscillations, which might be contribution to the coronal heating, as
seen in Appendix~B. The energy flow in the case of decayless
oscillations looks like the following: unknown driver $\rightarrow$
standing kink oscillations $\rightarrow$ quick damping
$\rightarrow$ plasma heating. Such process is continuing when the
unknown driver always operates, manifested as `decayless oscillations'.
However, the unknown driver is still an open issue, we will discuss
this in Section~5.3. The energy flux is estimated to
$\sim$7$-$9220~W~m$^{-2}$ with an average of 820~W~m$^{-2}$, the
median value is as low as 380~W~m$^{-2}$, suggesting that the lower
energy is dominant. The minimum energy flux (7~W~m$^{-2}$) taken by
the decayless kink oscillation corresponds to a longer oscillatory
period (169~s), but the amplitude and the magnetic field are rather
low, particularly the velocity amplitude is only 5.3~km~s$^{-1}$,
and the magnetic field perturbation is as weak as 0.07~G, as shown
in Table~\ref{tab_loop1} (slit S1, loop 12) and Figure~\ref{slt1}.
The same case of a lower energy flux (40~W~m$^{-2}$) has the longest
oscillatory period (185~s) and the weakest magnetic field strength
(4.1~G) in this statistic, as seen in Table~\ref{tab_loop1} (slit
S2, loop 6) and Figure~\ref{slt2}. In contrast, the maximum energy
flux (9220~W~m$^{-2}$) carried by the decayless kink oscillation
corresponds to a shorter oscillatory period (26~s), while the
amplitude and the magnetic field are stronger, particularly the
velocity amplitude can reach 118.8~km~s$^{-1}$, and the magnetic
field perturbation is as strong as 1.5~G, as seen in
Table~\ref{tab_loop1} (slit S1, loop 6) and Figure~\ref{slt1}. The
same case of a higher energy flux (3970~W~m$^{-2}$) also has the
shortest oscillatory period (11~s) and the strongest magnetic field
strength (25.2~G) in this statistical study, as seen in
Table~\ref{tab_loop1} (slit S3, loop 7) and Figure~\ref{slt3}. This
is consistent with the scatter plots in Figure~\ref{relat}, which
shows that the energy flux is strongly dependent on both the
oscillatory period and the magnetic field. However, the maximum
energy flux of 9220~W~m$^{-2}$ is still not enough to heat the AR
corona, such as 10$^4$~W~m$^{-2}$ \citep[cf.][]{Withbroe77}.
Moreover, the higher energy flux (i.e., $>$1000~W~m$^{-2}$) is not
dominant (Figure~\ref{sta2}), and the energy flux estimated by the
expression for the fast kink MHD wave (i.e., Equation~\ref{eq5}) is
always overestimated \citep{Goossens13}. In summary, the practical
energy carried by the decayless kink MHD wave is not sufficient for
heating the AR corona.

\subsection{Excitation of decayless kink oscillations}
Previous observations suggested that the decaying oscillation could
be impulsively excited by an external solar eruption that can be
clearly observed, such as solar flares and jets, EUV waves, or flux
ropes
\citep[e.g.,][]{Antolin11,Kumar13,Reeves20,Zhang20,Mandal22,Zhang22}.
Similarly, the decayless oscillation may be triggered by an
impulsive energy release \citep{Nakariakov21}. On the other hand,
the decayless oscillation often lasts for several wave periods or
even longer \citep{Tian12,Anfinogentov15}. Hence, the external
driver should be continuous to keep the decayless oscillation for a
long time, as seen in Appendix~B. Now, several theoretical models
have been proposed to illustrate the excitation/driver of decayless
oscillations, for instance, it might be triggered by the fast
magnetoacoustic wave train \citep{Liu11,Wang12}, or it could be
excited by the coronal rain caused by a catastrophic cooling process
\citep{Verwichte17}, or it may be a self oscillation that is driven
by the slipping interaction between oscillating loops and steady
external medium flows \citep{Nakariakov16,Nakariakov22}. However,
the triggered eruptions are difficult to detect, largely due to
their fine-scale structures, as shown in Figure~\ref{flar}, which
presents an overview of our observation.

Figure~\ref{flar}~(a) shows full-disk light curves in the X-rays
from 03:05~UT to 04:10~UT on 17 March 2022, which are recorded by
SolO/STIX and GOES, respectively. Here, the observed time of GOES
has been corrected to the SolO time stamps. There were not solar
flares appearing in the HRI$_{\rm EUV}$ observation during
03:18:00$-$04:02:51~UT, as indicated by the yellow shadow. We note
that a small flare erupted at about 03:11~UT, which occurred before
the HRI$_{\rm EUV}$ observation. The solar flare is estimated to be
a B7 class based on the STIX
counts\footnote{https://datacenter.stix.i4ds.net/view/plot/lightcurves}.
It was observed by the STIX channels at 4$-$10~keV and
15$-$25~keV, but it was impossible to determine the flare location
due to the weak counts. Herein, we plot the normalized EUV light
curve (normalized to their maximal intensity) integrated over the
active region with a FOV of $\sim$278$\times$278~Mm$^2$ observed by
SDO/AIA at 131~{\AA}, as shown by the cyan line in panel~(a). To
save the computing time, we use AIA binned (4$\times$4) images with
a time cadence of 120~s. Similar to STIX and GOES fluxes, the flare
peak also appears in the local EUV flux, indicating that the small
flare occurred inside the AR. To identify the flare source, we draw
the AIA~131~{\AA} image within a small FOV containing the AR, as
shown in Figure~\ref{flar}~(b). The small flare exhibits a bright
loop, as marked by the pink arrow. However, we should state that the
flare has disappeared before the beginning of decayless oscillation
observations. Moreover, most of oscillating loops are far away from
the flare source, especially for these oscillating loops in the R1
region. Therefore, the small flare has little impact on the
decayless oscillations in our study. Our results are similar to the
previous finding, for instance, the solar flare could increase the
oscillatory amplitude, but it had little effect on the nature and
periods of decayless oscillations \citep[cf.][]{Mandal21}.

Figure~\ref{flar}~(c) presents the FSI~304~{\AA} image at
03:50:00~UT during our observation. From which, we do not find any
coronal mass ejection (CME) or prominence eruptions, implying that
the decayless oscillation is impossible to be excited by the
precursor of a large-scale solar eruption, i.e., the solar flare,
the CME or the erupted prominence. Figure~\ref{flar}~(d) and (e)
shows the EUV images with sub-fields (marked by the cyan box in
panel~(c)) that including the AR measured in wavelengths of
HRI$_{\rm Ly\alpha}$~1216~{\AA} and FSI~304~{\AA}. Similarly, we do
not see any small-scale eruptions in these two EUV channels.
However, the two EUV images have lower spatial or temporal
resolutions, and the observational data is always discontinuous,
making it hard to synthesize the animation. Therefore, it is
impossible to exclude the possibility that the decayless
oscillations are excited by small-scale eruptions, particularly for
those small-scale eruptions with short durations, i.e., campfires
\citep{Berghmans21,Chen21}, ultraviolet bright points \citep{Li22c},
or fast repeating jets \citep{Chitta21,Hou21}, since they are
ubiquitous in the solar corona. On the other hand, the animation of
eui\_174.mp4 shows that a large number of thread-like structures
reveal continuous movements, which might be regarded as steady
flows. These steady flows could continuously interact with the
oscillating loops, causing decayless oscillations
\citep{Nakariakov16}. In a word, we can conclude that the decayless
kink oscillations at short periods could not be excited by
large-scale solar eruptions, but they might be triggered by
fine-scale solar eruptions, or they could be self oscillations
caused by slipping interactions between external flows and
oscillating loops.

\section{Summary}
Thanks to the high resolution observation from SolO/EUI at the
passband of HRI$_{\rm EUV}$~174~{\AA}, we present a statistical
study of short-period decayless oscillations of coronal loops in the
AR NOAA 12965, combining with observations from SolO/FSI, SolO/STIX,
SDO/AIA, and GOES, we discuss the excitation of observed decayless
kink oscillations, and their contribution to the coronal heating.
The main conclusions are summarized as following:

\begin{enumerate}

\item A total of 111 oscillating loops are extracted from the AR at the
HRI$_{\rm EUV}$~174~{\AA} passband. They all show transverse
oscillations without significant decaying, which could be identified
as decayless kink-mode MHD wave of AR coronal loops.

\item A strong correlation is found between oscillatory periods and loop
lengths, namely, the oscillatory period increases linearly with the
length of oscillating loops, confirming the interpretation of
standing kink-mode oscillations of coronal loops.

\item The oscillatory periods of decayless kink oscillations are dominated
by shorter periods, i.e., with a median period of 40~s. This is
benefitting from the high-temporal ($\sim$3~s) and high-spatial
($\sim$135~km~pixel$^{-1}$) resolution observation measured by
SolO/EUI at the channel of HRI$_{\rm EUV}$~174~{\AA}.

\item The displacement amplitudes of decayless kink oscillations are
rather low, but their velocity amplitudes could be as large as
118.8~km~s$^{-1}$. Thus, these decayless kink oscillations could not
be called as low-amplitude oscillations, but they are regarded as
small-scale oscillations.

\item The energy fluxes carried by decayless kink oscillations are
estimated to be 9$-$9220~W~m$^{-2}$, after considering the magnetic
energy. They are dominated by energy fluxes below 1000~W~m$^{-2}$,
for instance, the averaged energy flux is 820~W~m$^{-2}$, while the
median value is only 380~W~m$^{-2}$. So, the decayless kink-mode
wave could not be efficiently heating the AR corona, although they
might be persistently powered and ongoing dissipating energies that
are transferred to the ambient coronal plasmas.

\item The short-period decayless oscillation in the kink mode could be a
self-oscillation system excited by the slipping interaction between
the external flows and oscillating loops. However, we could not
excluded the possibility that it was triggered by a fine-scale solar
eruption, mainly due to the lack of higher spatial resolution
observations in the chromosphere.

\end{enumerate}

\acknowledgments We thank the referee for his/her valuable comments.
The author would like to thank the EUI team members
for their discussions about the data analysis. This work is funded by the National Key R\&D Program of
China 2021YFA1600502 (2021YFA1600500), NSFC under grants 11973092, U1931138, 12073081, 11790302, as well
as CAS Strategic Pioneer Program on Space Science, Grant No.
XDA15052200, and XDA15320301. D.~Li is also supported by the Surface
Project of Jiangsu Province (BK20211402). D.~M.~Long is grateful to
the Science Technology and Facilities Council for the award of an
Ernest Rutherford Fellowship (ST/R003246/1). Solar Orbiter is a
space mission of international collaboration between ESA and NASA,
operated by ESA. The EUI instrument was built by CSL, IAS, MPS,
MSSL/UCL, PMOD/WRC, ROB, LCF/IO with funding from the Belgian
Federal Science Policy Office (BELSPO/PRODEX PEA under contract
numbers 4000134088, 4000112292, 4000117262, and 4000134474); the
Centre National d'Etudes Spatiales (CNES); the UK Space Agency
(UKSA); the Bundesministerium f\"{u}r Wirtschaft und Energie (BMWi)
through the Deutsches Zentrum f\"{u}r Luft- und Raumfahrt (DLR); and
the Swiss Space Office (SSO). The STIX instrument is an
international collaboration between Switzerland, Poland, France,
Czech Republic, Germany, Austria, Ireland, and Italy.

\clearpage

\begin{table*}[ht]
\addtolength{\tabcolsep}{2pt} \centering \caption{Observed Instruments used
in this study.} \label{tab_instr}
\begin{tabular}{cccccc}
\hline \hline
Telescope  &  Wavelength   &  Time cadence (s)  & Pixel scale (km)    &  Channel    &   Time (UT)            \\
\hline
SolO/EUI    &  174~{\AA}    &   $\sim$3        & $\sim$135        &   EUV         &   03:18:00$-$04:02:51   \\
            &  1216~{\AA}   &   $\sim$5        & $\sim$282        &   Ly$\alpha$  &   03:18:00$-$03:46:55   \\
\hline
SolO/FSI   &  304~{\AA}     &   $\sim$60       & $\sim$1219       &   EUV         &   03:50:00$-$04:03:00   \\
\hline
SDO/AIA    &  171~{\AA}     &     12           & $\sim$435        &   EUV         &       --                \\
           &  131~{\AA}     &     120          & $\sim$1740       &   EUV         &       --                \\
\hline
           &  4$-$10~keV      &  $\sim$4       &     --           &  SXR          &                         \\
SolO/STIX  &  10$-$15~keV     &  $\sim$4       &     --           &  SXR          &   02:49:49$-$06:09:57   \\
           &  15$-$25~keV     &  $\sim$4       &     --           &  SXR/HXR      &                         \\
\hline
GOES       &  1$-$8~{\AA}     &  $\sim$1       &     --           &  SXR          &       --                \\
\hline \hline
\end{tabular}
\end{table*}

\begin{center}
\tabcolsep 3pt  
\begin{longtable}{cccccccccc}
   \caption{Key parameters measured in oscillating loops.}
   \label{tab_loop1} \\
   \toprule
Slit  &  Loops & $L$ (Mm) & $P$ (s) &  $A_m$ (km) &  $v_{m}$ (km~s$^{-1}$)  & $c_{k}$ (km~s$^{-1}$) &  $B$ (G) & $b$ (G) & $E$ (W~m$^{-2}$)  \\
\hline
      &  1   &  18.4    &   76    &   275  &  22.6   & 480  & 6.0    & 0.28  & 180      \\
      &  2   &  14.9    &   56    &   181  &  20.2   & 530  & 6.6    & 0.25  & 160      \\
      &  3   &  11.2    &   25    &   156  &  39.0   & 890  & 11.2   & 0.49  & 990     \\
      &  4   &  10.9    &   24    &   386  &  97.5   & 880  & 11.0   & 1.22  & 6100    \\
      &  5   &  10.7    &   22    &   126  &  35.5   & 960  & 12.1   & 0.45  & 880     \\
      &  6   &  11.5    &   26    &   488  &  118.8  & 890  & 11.2   & 1.50  & 9220    \\
      &  7   &  24.9    &   145   &   148  &  6.4    & 340  & 4.3    & 0.08  & 10       \\
S1    &  8   &  16.4    &   62    &   259  &  26.2   & 530  & 6.6    & 0.33  & 270      \\
      &  9   &  11.7    &   26    &   164  &  39.5   & 900  & 11.3   & 0.50  & 1020     \\
      &  10  &  13.5    &   34    &   177  &  33.2   & 800  & 10.1   & 0.42  & 650     \\
      &  11  &  12.4    &   33    &   249  &  47.1   & 750  & 9.4    & 0.59  & 1210     \\
      &  12  &  29.2    &   169   &   144  &  5.3    & 350  & 4.3    & 0.07  &  7       \\
      &  13  &  22.9    &   113   &   421  &  23.4   & 410  & 5.1    & 0.29  & 160      \\
\hline
      &  1   &  12.8    &   25    &   98   &  24.7  & 1020 & 12.9  & 0.31  & 460      \\
      &  2   &  15.7    &   47    &   254  &  33.6  & 660  & 8.3   & 0.42  & 550      \\
S2    &  3   &  17.1    &   70    &   350  &  31.5  & 490  & 6.1   & 0.40  & 350       \\
      &  4   &  12.3    &   26    &   152  &  36.8  & 950  & 11.9  & 0.46  & 940      \\
      &  5   &  11.8    &   20    &   164  &  52.4  & 1200 & 15.1  & 0.66  & 2410      \\
      &  6   &  30.2    &   185   &   358  &  12.1  & 330  & 4.1   & 0.15  & 40        \\
\hline
      &  1   &  11.2    &   26    &   132  &  32.0   & 860  & 10.8   & 0.40  & 650  \\
      &  2   &  22.3    &   113   &   352  &  19.6   & 390  & 4.9    & 0.25  & 110   \\
      &  3   &  13.3    &   32    &   124  &  24.7   & 840  & 10.6   & 0.31  & 380   \\
      &  4   &  16.4    &   62    &   147  &  14.9   & 530  & 6.7    & 0.19  & 90   \\
S3    &  5   &  16.5    &   62    &   131  &  13.2   & 530  & 6.7    & 0.17  & 70   \\
      &  6   &  14.9    &   40    &   255  &  39.6   & 740  & 9.3    & 0.50  & 850  \\
      &  7   &  10.5    &   11    &   91   &  52.0   & 1910 & 25.2   & 0.69  & 3970  \\
      &  8   &  15.4    &   44    &   52   &  7.4    & 700  & 8.8    & 0.09  &  30   \\
      &  9   &  15.9    &   54    &   122  &  14.2   & 590  & 7.4    & 0.18  &  90  \\
      &  10  &  15.8    &   53    &   263  &  31.2   & 600  & 7.5    & 0.39  &  430  \\
\hline
      &  1   &  11.9    &   25    &   168  &  41.5   & 940  & 11.8   & 0.52  & 1180    \\
      &  2   &  16.0    &   49    &   439  &  56.2   & 650  & 8.2    & 0.71  & 1510    \\
      &  3   &  21.9    &   115   &   148  &  8.1    & 380  & 4.8    & 0.10  & 20      \\
S4    &  4   &  16.1    &   49    &   118  &  15.1   & 650  & 8.2    & 0.19  & 110     \\
      &  5   &  13.3    &   37    &   140  &  23.8   & 720  & 9.0    & 0.30  & 300     \\
      &  6   &  14.5    &   45    &   237  &  33.2   & 640  & 8.1    & 0.42  & 520     \\
\hline
      &  1   &  13.8    &   30    &   185  &  39.0   & 930  & 11.7   & 0.49  & 1030     \\
      &  2   &  11.9    &   24    &   201  &  53.0   & 1000 & 12.5   & 0.67  & 2050     \\
      &  3   &  15.7    &   50    &   239  &  30.1   & 630  & 7.9    & 0.38  & 420      \\
S5    &  4   &  13.7    &   30    &   157  &  33.4   & 920  & 11.6   & 0.42  & 750      \\
      &  5   &  10.8    &   21    &   186  &  55.1   & 1020 & 12.8   & 0.69  & 2260     \\
      &  6   &  14.0    &   36    &   189  &  33.5   & 790  & 9.9    & 0.42  & 640      \\
      &  7   &  12.0    &   26    &   141  &  33.4   & 900  & 11.4   & 0.42  & 740      \\

\hline
      &  1   &  22.9    &   120   &   367  &  19.2   & 380  & 4.8   & 0.24  & 100   \\
      &  2   &  16.8    &   66    &   180  &  17.1   & 510  & 6.4   & 0.21  & 110   \\
S6    &  3   &  12.5    &   27    &   343  &  81.0   & 940  & 11.8  & 1.0   & 4500  \\
      &  4   &  16.6    &   54    &   138  &  16.0   & 610  & 7.7   & 0.20  & 120   \\
      &  5   &  16.7    &   57    &   153  &  16.9   & 590  & 7.4   & 0.21  & 120   \\
\hline
      &  1   &  13.5    &   37    &   200  &  33.5  & 720  & 9.0   & 0.42  & 590     \\
      &  2   &  15.3    &   49    &   219  &  28.0  & 630  & 7.9   & 0.35  & 360     \\
      &  3   &  21.2    &   98    &   126  &  8.1   & 430  & 5.4   & 0.10  & 20      \\
S7    &  4   &  15.1    &   44    &   176  &  25.2  & 690  & 8.7   & 0.32  & 320     \\
      &  5   &  10.5    &   23    &   83   &  22.7  & 910  & 11.5  & 0.29  & 350    \\
      &  6   &  15.2    &   54    &   334  &  39.1  & 570  & 7.1   & 0.49  & 640    \\
      &  7   &  14.9    &   53    &   164  &  19.2  & 560  & 7.0   & 0.24  & 150    \\
      &  8   &  24.6    &   135   &   385  &  18.0  & 360  & 4.6   & 0.23  & 90     \\
\hline
      &  1   &  18.8    &   77    &   324  &  26.5  & 490  & 6.2   & 0.33  & 250    \\
      &  2   &  13.9    &   39    &   433  &  69.7  & 710  & 8.9   & 0.88  & 2520   \\
S8    &  3   &  13.6    &   38    &   280  &  46.1  & 720  & 9.0   & 0.58  & 1110    \\
      &  4   &  13.9    &   40    &   211  &  32.8  & 690  & 8.7   & 0.41  & 510     \\
      &  5   &  14.1    &   40    &   201  &  31.6  & 710  & 8.9   & 0.40  & 510     \\
      &  6   &  16.5    &   58    &   326  &  35.4  & 570  & 7.2   & 0.44  & 520     \\
\hline
      &  1   &  14.4    &   42    &   139  &  20.9   & 690  & 8.6    & 0.26  & 220   \\
      &  2   &  13.2    &   27    &   129  &  30.5   & 990  & 12.4   & 0.38  & 670  \\
      &  3   &  12.4    &   17    &   98   &  36.6   & 1480 & 18.5   & 0.46  & 1450  \\
      &  4   &  13.0    &   23    &   77   &  20.9   & 1130 & 14.2   & 0.26  & 360   \\
      &  5   &  13.0    &   23    &   152  &  41.2   & 1130 & 14.2   & 0.52  & 1400  \\
      &  6   &  14.0    &   38    &   157  &  25.7   & 730  & 9.2    & 0.32  & 350   \\
      &  7   &  12.5    &   17    &   131  &  48.0   & 1450 & 18.2   & 0.60  & 2440  \\
      &  8   &  13.6    &   26    &   162  &  39.3   & 1050 & 13.1   & 0.49  & 1180  \\
S9    &  9   &  12.8    &   23    &   104  &  28.3   & 1110 & 13.9   & 0.36  & 650  \\
      &  10  &  14.2    &   37    &   201  &  33.9   & 760  & 9.6    & 0.43  & 640  \\
      &  11  &  18.2    &   73    &   450  &  38.8   & 500  & 6.3    & 0.49  & 550   \\
      &  12  &  13.1    &   25    &   151  &  38.6   & 1070 & 13.4   & 0.48  & 1160  \\
      &  13  &  14.8    &   39    &   115  &  18.5   & 760  & 9.5    & 0.23  & 190   \\
      &  14  &  15.5    &   49    &   132  &  17.0   & 630  & 8.0    & 0.21  & 130   \\
      &  15  &  14.6    &   41    &   92   &  14.0   & 710  & 8.9    & 0.18  & 100   \\
      &  16  &  13.2    &   26    &   174  &  42.7   & 1030 & 12.9   & 0.54  & 1370  \\
      &  17  &  21.8    &   113   &   119  &  6.6    & 390  & 4.9    & 0.08  & 10    \\
      &  18  &  14.4    &   41    &   293  &  45.5   & 710  & 8.9    & 0.57  & 1080  \\
\hline
      &  1   &  13.6    &   30    &   84   &  17.8   & 920  & 11.5   & 0.22  &  210  \\
      &  2   &  11.0    &   13    &   133  &  63.1   & 1670 & 21.0   & 0.79  &  4870 \\
      &  3   &  14.5    &   44    &   143  &  20.6   & 670  & 8.4    & 0.26  &  210  \\
      &  4   &  13.8    &   32    &   221  &  43.1   & 860  & 10.8   & 0.54  &  1160 \\
      &  5   &  13.3    &   29    &   398  &  86.4   & 920  & 11.5   & 1.09  &  5020 \\
      &  6   &  15.7    &   50    &   175  &  22.2   & 640  & 8.0    & 0.28  &  230  \\
S10   &  7   &  16.1    &   53    &   102  &  12.2   & 610  & 7.6    & 0.15  &  70   \\
      &  8   &  15.2    &   47    &   123  &  16.2   & 640  & 8.1    & 0.20  &  120  \\
      &  9   &  16.4    &   55    &   108  &  12.3   & 590  & 7.4    & 0.15  &  70   \\
      &  10  &  17.1    &   62    &   87   &  8.8    & 550  & 6.9    & 0.11  &  30   \\
      &  11  &  12.7    &   22    &   88   &  25.6   & 1170 & 14.7   & 0.32  &  560  \\
      &  12  &  12.0    &   17    &   68   &  25.5   & 1430 & 18.1   & 0.32  &  680  \\
      &  13  &  12.8    &   21    &   90   &  26.9   & 1220 & 15.4   & 0.34  &  640  \\
      &  14  &  10.9    &   15    &   51   &  21.7   & 1490 & 18.7   & 0.27  &  520  \\
      &  15  &  11.5    &   18    &   73   &  25.7   & 1290 & 16.2   & 0.32  &  620  \\
\hline
      &  1   &  17.1    &   66    &  208   &  19.9   & 520  & 6.5    & 0.25  & 150   \\
      &  2   &  16.0    &   53    &  169   &  20.1   & 610  & 7.6    & 0.25  & 180   \\
S11   &  3   &  15.3    &   48    &  174   &  23.0   & 640  & 8.1    & 0.29  & 250   \\
      &  4   &  14.5    &   43    &  112   &  16.6   & 680  & 8.6    & 0.21  & 140   \\
\hline
      &  1   &  22.0    &   112   &   305  &  17.1   & 400  & 4.9    & 0.22  &  80  \\
      &  2   &  12.6    &   22    &   71   &  20.2   & 1150 & 14.5   & 0.25  &  350  \\
S12   &  3   &  12.5    &   22    &   70   &  20.3   & 1150 & 14.4   & 0.26  &  350  \\
      &  4   &  12.8    &   23    &   73   &  20.3   & 1130 & 14.2   & 0.25  &  340  \\
      &  5   &  15.5    &   49    &   74   &  9.5    & 640  & 8.0    & 0.12  &  40  \\
      &  6   &  15.8    &   50    &   111  &  14.0   & 630  & 7.9    & 0.18  &  90  \\
\hline
      &  1   &  16.3    &   62    &   149  &  15.1   & 530  & 6.6    & 0.19  & 90   \\
      &  2   &  10.9    &   20    &   80   &  25.6   & 1120 & 14.0   & 0.32  & 530  \\
      &  3   &  15.6    &   58    &   170  &  18.4   & 540  & 6.8    & 0.23  & 130   \\
S13   &  4   &  16.0    &   60    &   132  &  13.8   & 530  & 6.7    & 0.17  & 70    \\
      &  5   &  11.3    &   23    &   123  &  34.2   & 990  & 12.5   & 0.43  & 850   \\
      &  6   &  11.2    &   23    &   146  &  40.6   & 990  & 12.5   & 0.51  & 1200  \\
      &  7   &  27.1    &   165   &   249  &  9.5    & 330  & 4.1    & 0.12  &  20   \\
  \bottomrule
\end{longtable}
\end{center}

\begin{table}
\caption{Statistical results of key parameters measured in 111 oscillating loops.}
\label{tab_loop2}      
\tabcolsep 15pt        
\begin{tabular}{ccccccccc}
\toprule
                        &  Minimum & Maximum &  Median &  Average & Standard deviation \\
\hline
$L$ (Mm)                &    10.5  &  30.2   &  14.4 &  15.1     &  3.8               \\
$P$ (s)                 &    11    &  185    &  40   &  49       &  34                \\
$A_{m}$ (km)            &    51    &  488    &  156  &  186      &  99                \\
$v_{m}$ (km~s$^{-1}$)   &    5.3   &  118.8  &  25.6 &  29.8     &  18.5              \\
$c_{k}$ (km~s$^{-1}$)  &    330   &  1910   &  710  &  780      &  300               \\
$B$ (G)                 &    4.1   &  25.2   &  8.9  &  9.8      &  3.8               \\
$b$ (G)                 &    0.07  &  1.5    &  0.32 &  0.37     &  0.23              \\
$E$ (W~m$^{-2}$)        &    7     &  9220   &  380  &  820      &  1340              \\
\bottomrule
\end{tabular}
\end{table}

\begin{figure}
\centering
\includegraphics[width=0.5\linewidth,clip=]{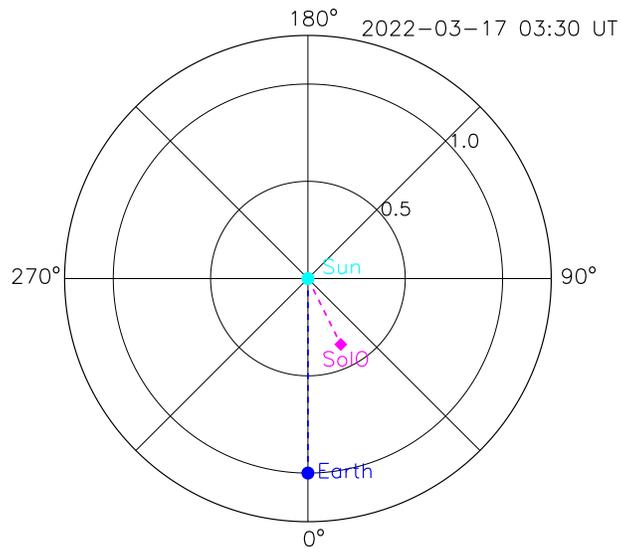}
\caption{Sketch plot of the location of SolO with respect to the
Earth at 03:30~UT on 17 March 2022, in Heliocentric-Earth equatorial
coordinates. \label{location}}
\end{figure}

\begin{figure}
\centering
\includegraphics[width=0.8\linewidth,clip=]{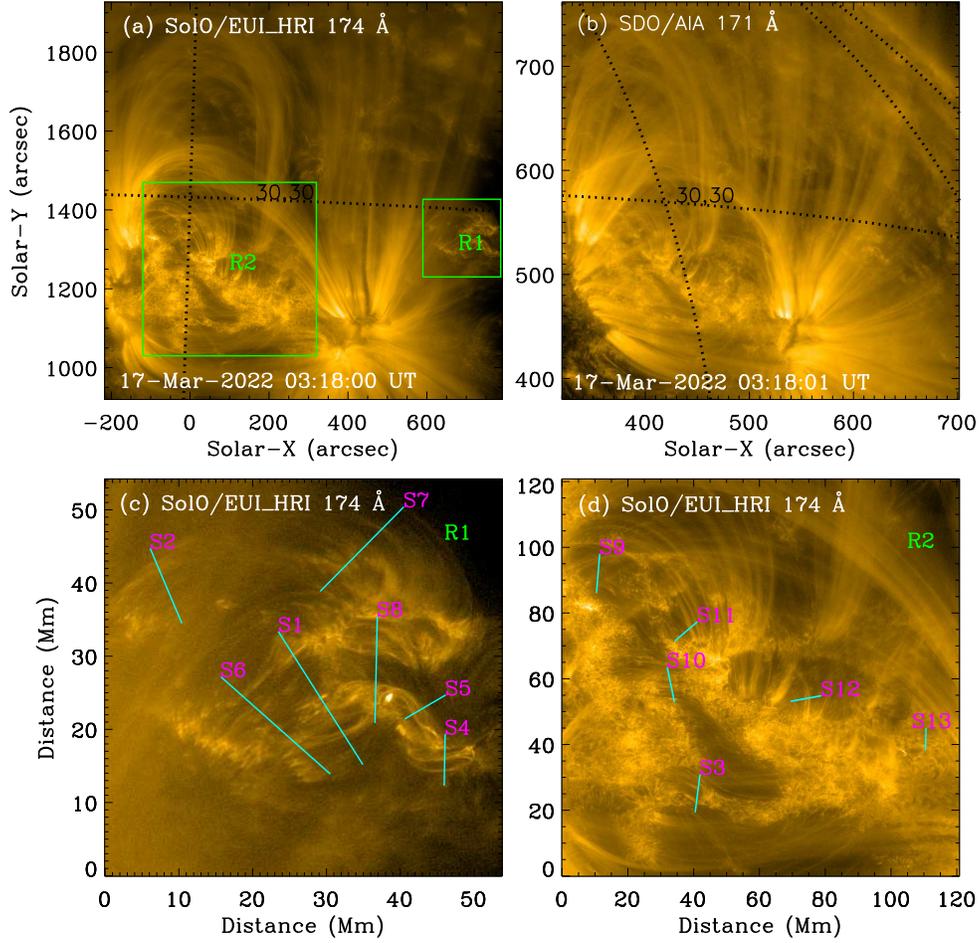}
\caption{Upper: Overview of coronal loops on 17 March 2022 as seen
by SolO/EUI at HRI$_{\rm EUV}$~174~{\AA} (a), and SDO/AIA~171~{\AA}
(b). The dotted lines mark the latitude-longitude grids. Bottom:
Sub-fields corresponding to two green boxes (R1 and R2) marked in
panel~(a). The cyan lines outline thirteen slits that are crossed to
coronal loops. An animation provided in the online version shows the
temporal evolution of coronal loops in two zoomed images (R1 and R2)
at HRI$_{\rm EUV}$~174~{\AA}. \label{img}}
\end{figure}

\begin{figure}
\centering
\includegraphics[width=0.8\linewidth,clip=]{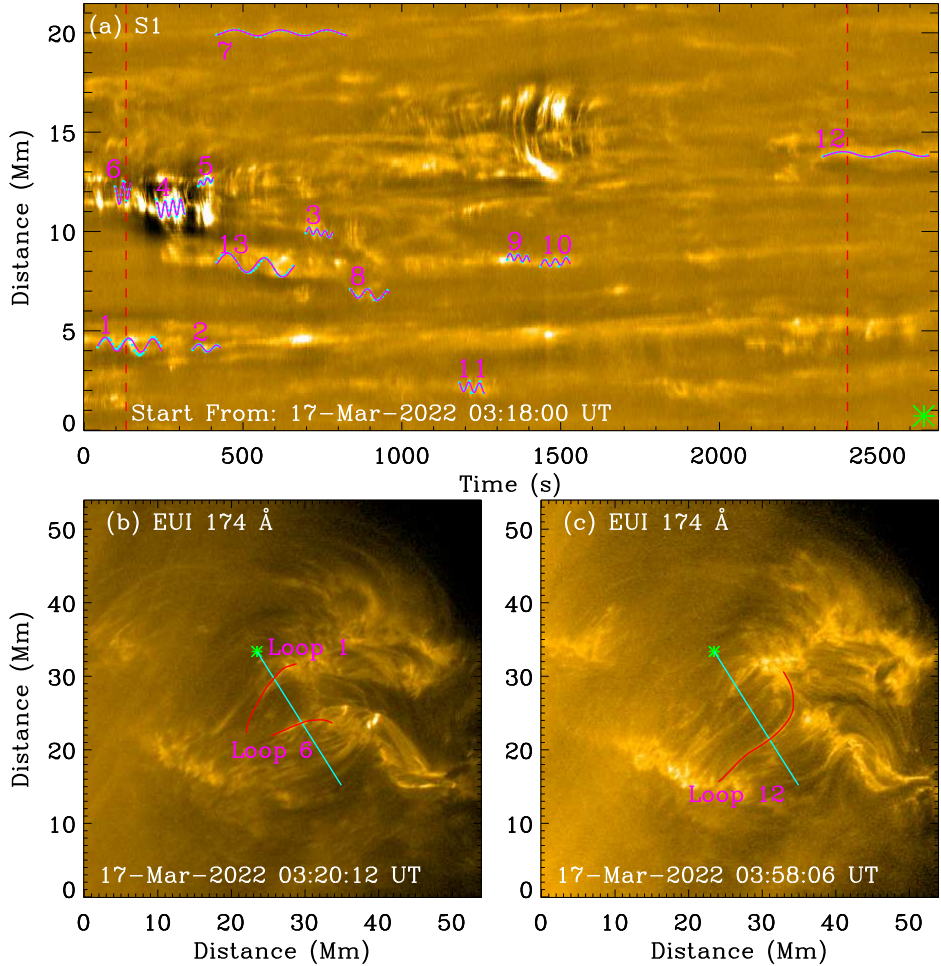}
\caption{Panel~(a): The time-distance map made from slit~S1 in
region~R1. The cyan circles indicate the fitting profile positions
of coronal loop oscillations, and the magenta curves represent their
best fitting results. Two vertical dashed lines outline the time in
panels~(b) \& (c), respectively. Panels~(b) \& (c): Two EUV images
show the interested coronal loops (1, 6 and 12), as marked by the
red curves. The cyan line indicates the slit position, and the green
symbol (`$\ast$') denotes the starting point of the
time-distance analysis. \label{slt1}}
\end{figure}

\begin{figure}
\centering
\includegraphics[width=0.8\linewidth,clip=]{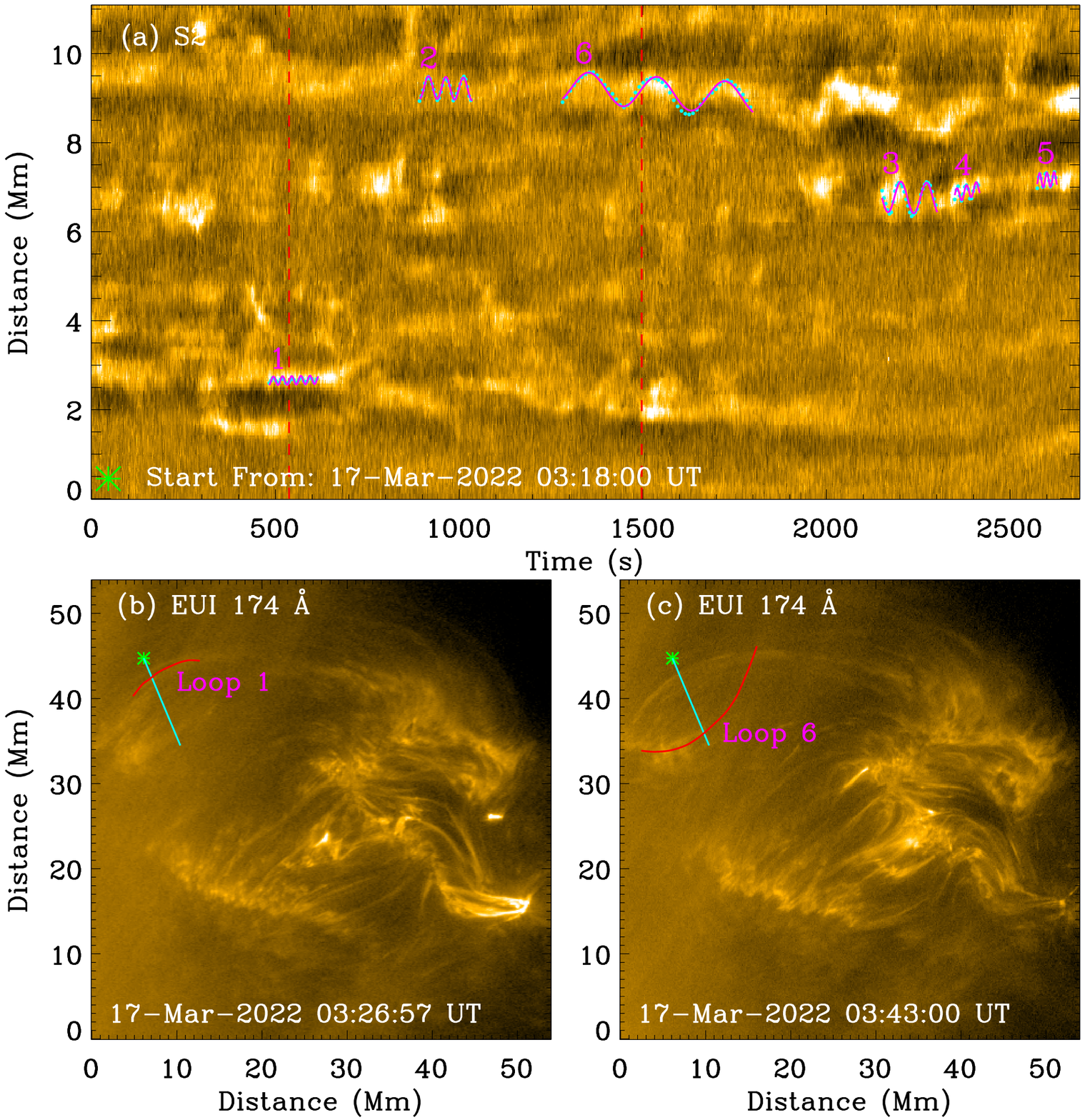}
\caption{Similar to Figure~\ref{slt1} but the analysis is made for
slit~S2 in region~R1. \label{slt2}}
\end{figure}

\begin{figure}
\centering
\includegraphics[width=0.8\linewidth,clip=]{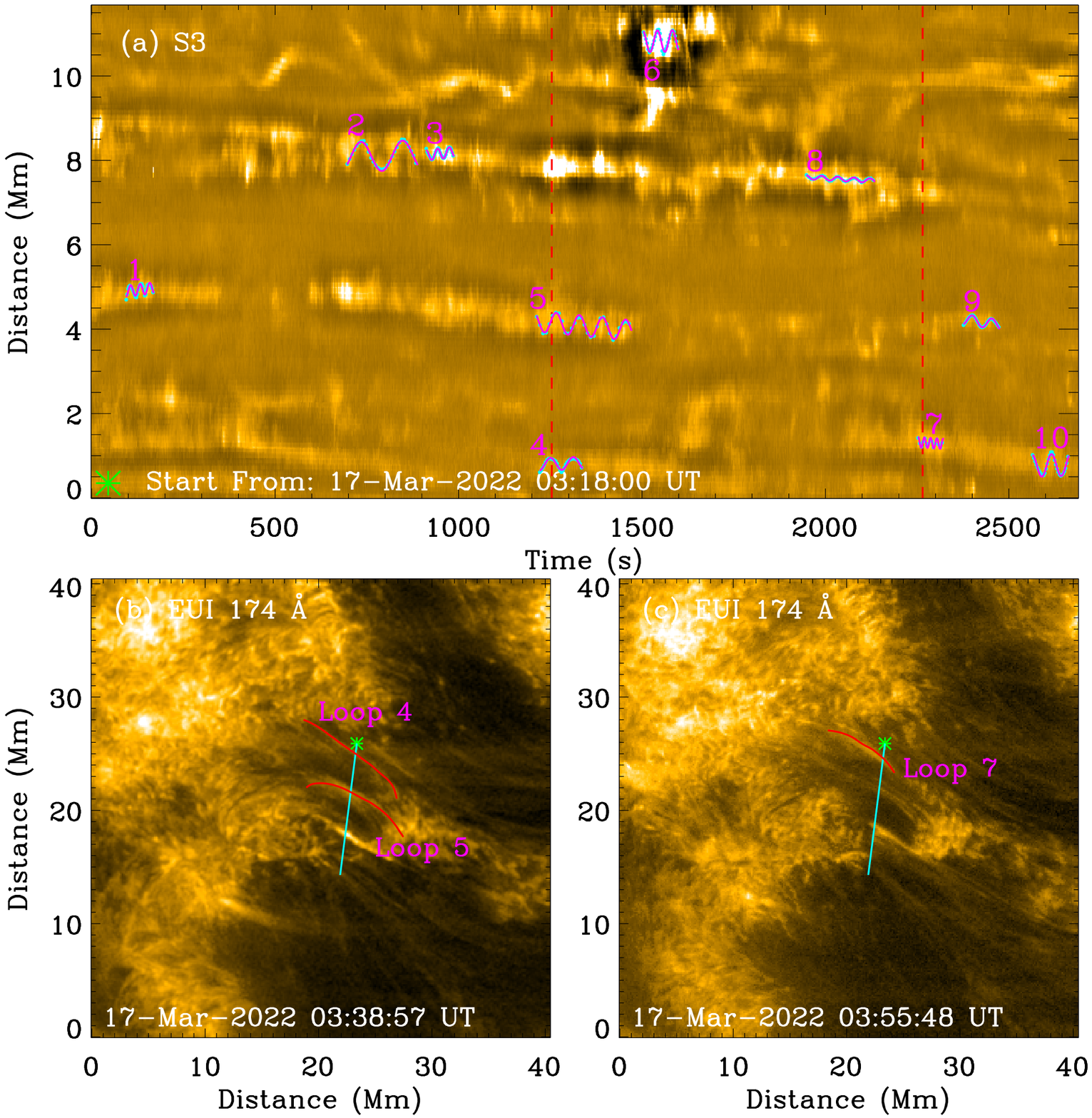}
\caption{Similar to Figure~\ref{slt1} but the analysis is made for
slit~S3 in region~R2. \label{slt3}}
\end{figure}

\begin{figure}
\centering
\includegraphics[width=0.8\linewidth,clip=]{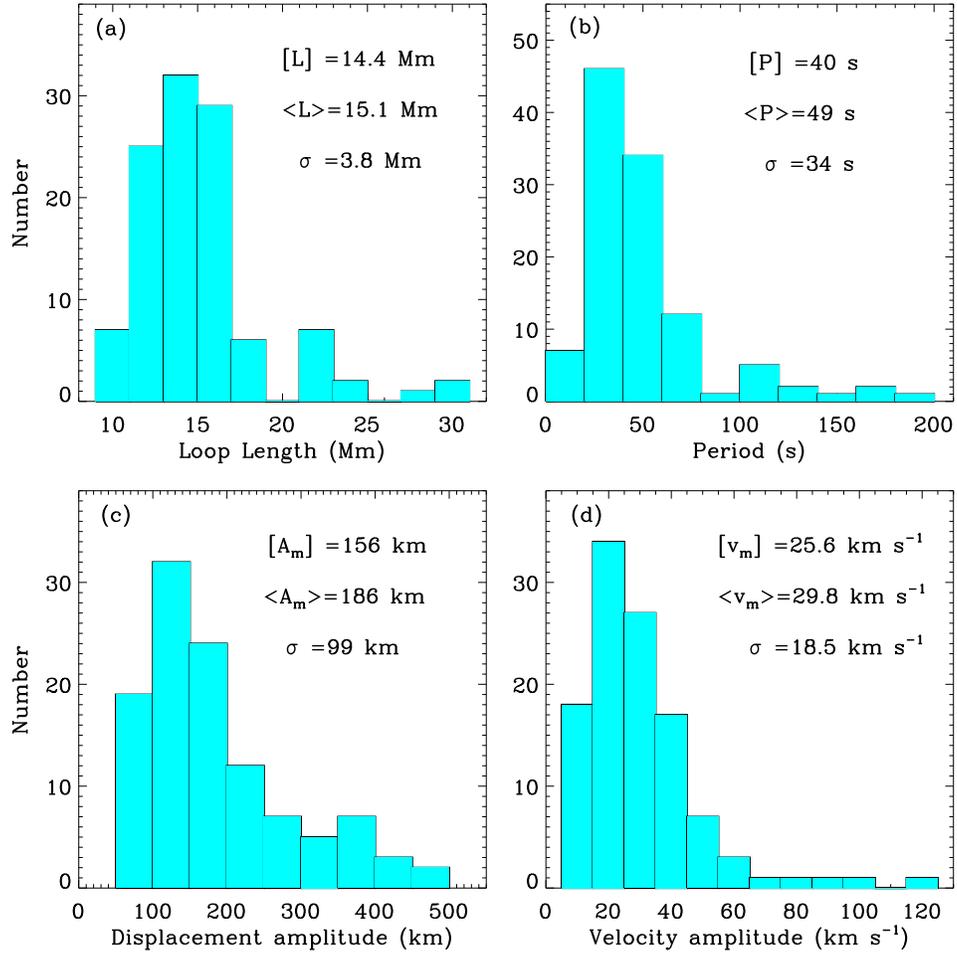}
\caption{Distributions of the main oscillatory parameters: loop
length (a), oscillatory period (b), displacement amplitude (c),
and velocity amplitude (d). The median ($[~~]$) values, average
($<>$) values and standard deviations ($\sigma$) for the
corresponding parameters are labeled in each panel. \label{sta1}}
\end{figure}

\begin{figure}
\centering
\includegraphics[width=0.8\linewidth,clip=]{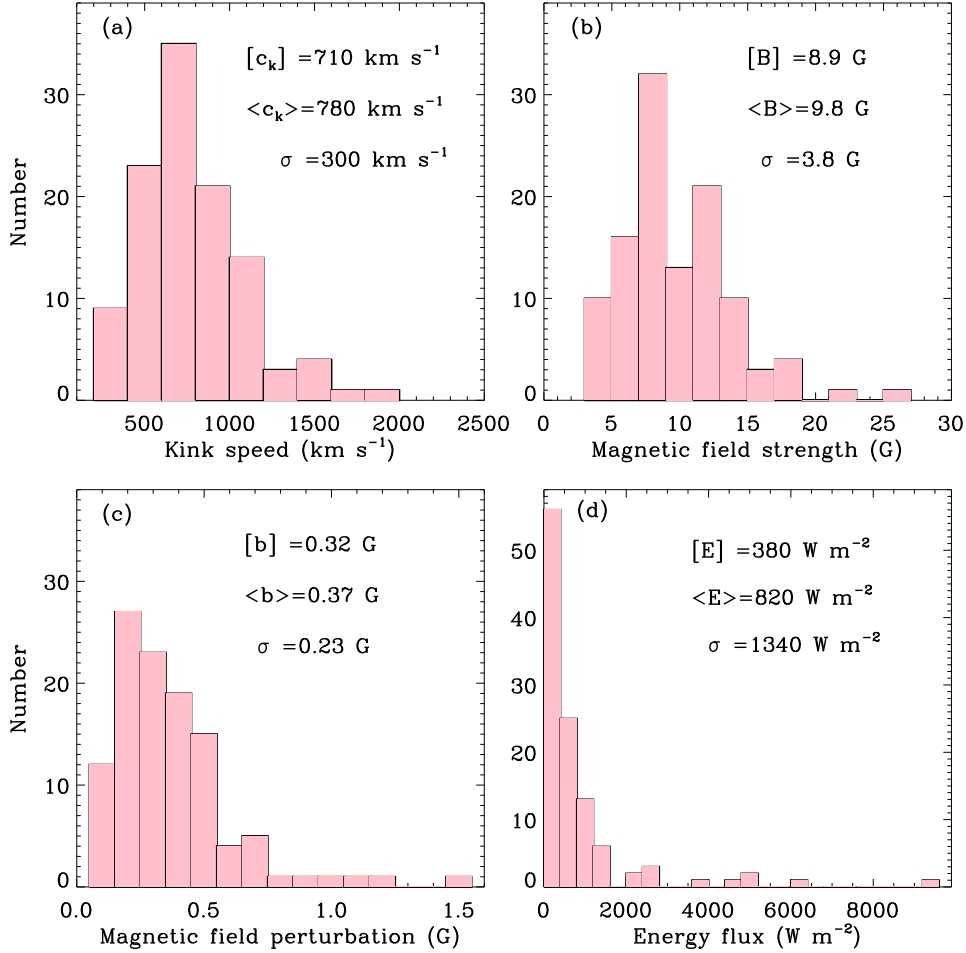}
\caption{Distributions of the key physical parameters from coronal
loop oscillations: kink speed (a), magnetic field strength (b),
magnetic field perturbation (c), and energy flux (d). The median
($[~~]$) values, average ($<>$) values and standard deviations
($\sigma$) for the corresponding parameters are labeled in each
panel. \label{sta2}}
\end{figure}

\begin{figure}
\centering
\includegraphics[width=0.6\linewidth,clip=]{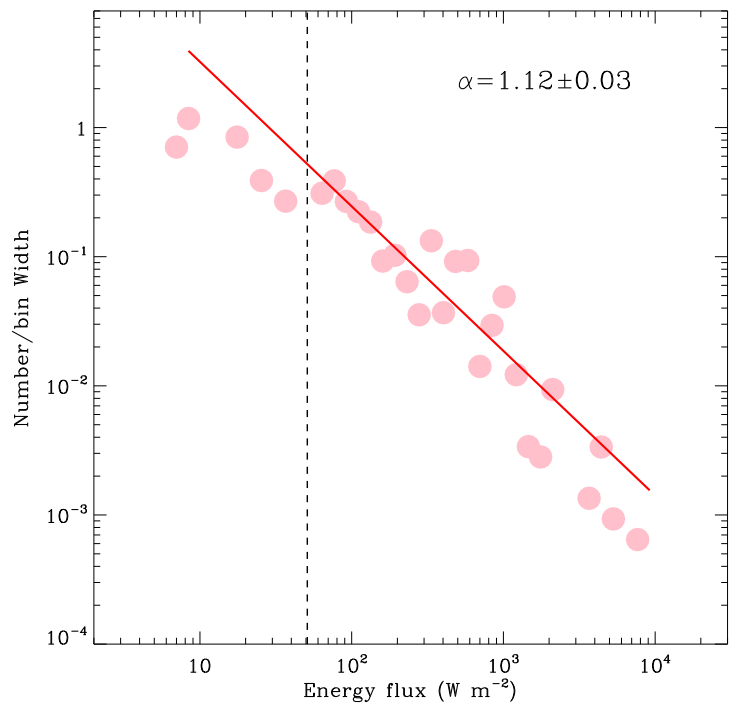}
\caption{The frequency distribution as a function of the energy flux
in log-log space. The power-law index ($\alpha$) is estimated by the
maximum likelihood method. The vertical line indicates the low
energy cutoff, and the red line represents the power-law model.
\label{sta3}}
\end{figure}

\begin{figure}
\centering
\includegraphics[width=0.8\linewidth,clip=]{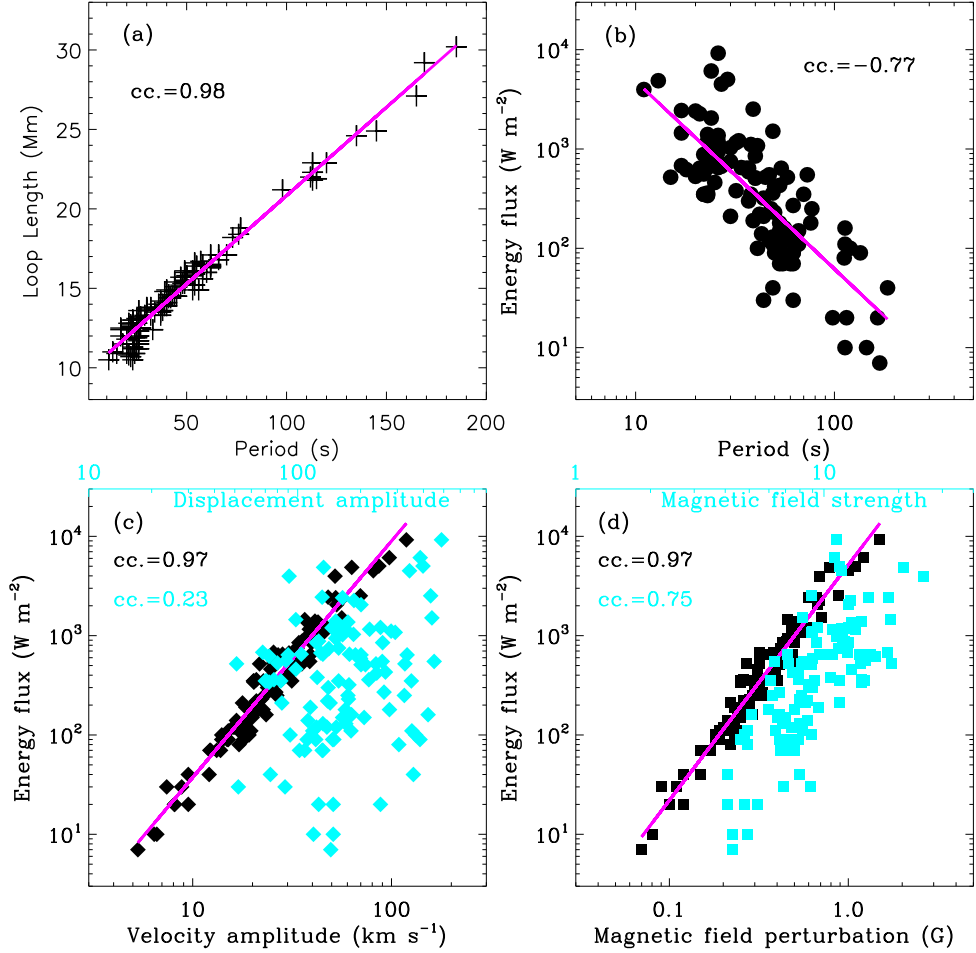}
\caption{Scatter plots between two key parameters of coronal loop
oscillations: oscillatory periods vs. loop lengths (a), oscillatory
periods vs. energy flux (b), velocity amplitudes (black) and
displacement amplitude (cyan) vs. energy flux (c), magnetic field
perturbation (black) and magnetic field strength (cyan) vs. energy
flux (d). The Pearson correlation coefficients (cc.) are also
provided. The magenta line indicates a linear fit in each panel.
\label{relat}}
\end{figure}

\begin{figure}
\centering
\includegraphics[width=0.8\linewidth,clip=]{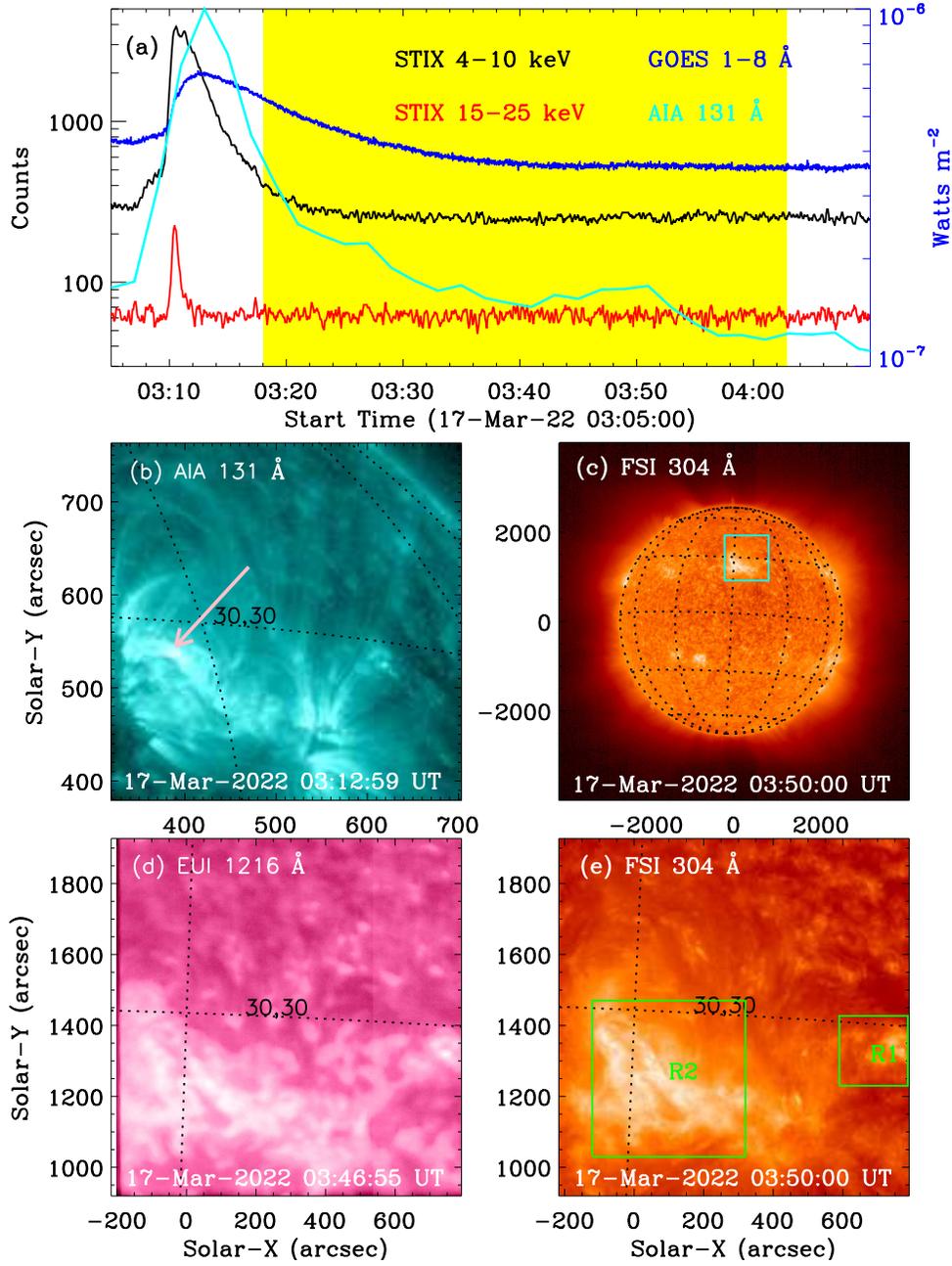}
\caption{Panel~(a): Multi-wavelength time series during
03:05$-$04:10~UT on 17 March 2022, namely full-disk light curves
recorded by GOES 1$-$8~{\AA}(blue), STIX~4$-$10~keV (black),
10$-$15~keV (red) and 15$-$25~keV (green), and the local flux
integrated over the active region in AIA~131~{\AA} as normalization
by the maximal intensity. The yellow shadow indicates the
observational duration of SolO/EUI~174~{\AA}. Panel~(b):
SDO/AIA~131~{\AA} image shows the active region used to integrate
the local flux, the pink arrow indicates the flare site. Panel~(c):
SolO/FSI~304 map shows the full Sun, the cyan box outlines the
interested active region. The dotted lines mark the
latitude-longitude grids. Panels~(d) and (e): EUV images show the
interested active region (cyan box in panel~(c)), observed by
SolO/EUI at Ly$\alpha$~1216~{\AA}, and SolO/FSI at 304~{\AA},
respectively. The green boxes outline the same sub-fields in
Figure~\ref{img} \label{flar}}
\end{figure}

\clearpage
\appendix
\renewcommand\thefigure{\Alph{section}\arabic{figure}}
\section{Zoomed time-distance images}
\setcounter{figure}{0}

Figures~\ref{slt1}$-$\ref{slt3} present time-distance maps along
longer slits during the whole observational time, i.e., from
03:18:00~UT to 04:02:51~UT. Thus, some decayless oscillations could
be not very obvious, mainly due to their small-scale amplitudes
overplotted on the large FOV map. Actually, the small-scale
oscillation with a short period is identified from the zoomed
time-distance map one by one, as shown in
Figures~\ref{fc_s1}$-$\ref{fc_s3}. Figure~\ref{fc_s1} displays two
zoomed time-distance maps along slit~1 in region~R1, corresponding
to the oscillating loops 1 (a) and 12 (b), respectively. Those cyan
circles mark the central positions of the oscillating loops. Similar
to previous studies \citep{Anfinogentov13,Anfinogentov15}, the loop
centers are not determined by Gaussian fitting. This is because that
another coronal loop without any oscillatory signature is
overlapping on the oscillating loop, as indicated by two pink
arrows. On the other hand, the loop centers of some oscillating
loops are difficult to identify, due to the overlap of multiple
loops. In such case, the loop edges are used to study their
oscillations according to the assumption of constant cross-sectional
loops \citep[cf.][]{Williams21,Gao22}, as shown in
Figure~\ref{fc_s2}~(b) and Figure~\ref{fc_s3}~(a).

\begin{figure}
\centering
\includegraphics[width=0.8\linewidth,clip=]{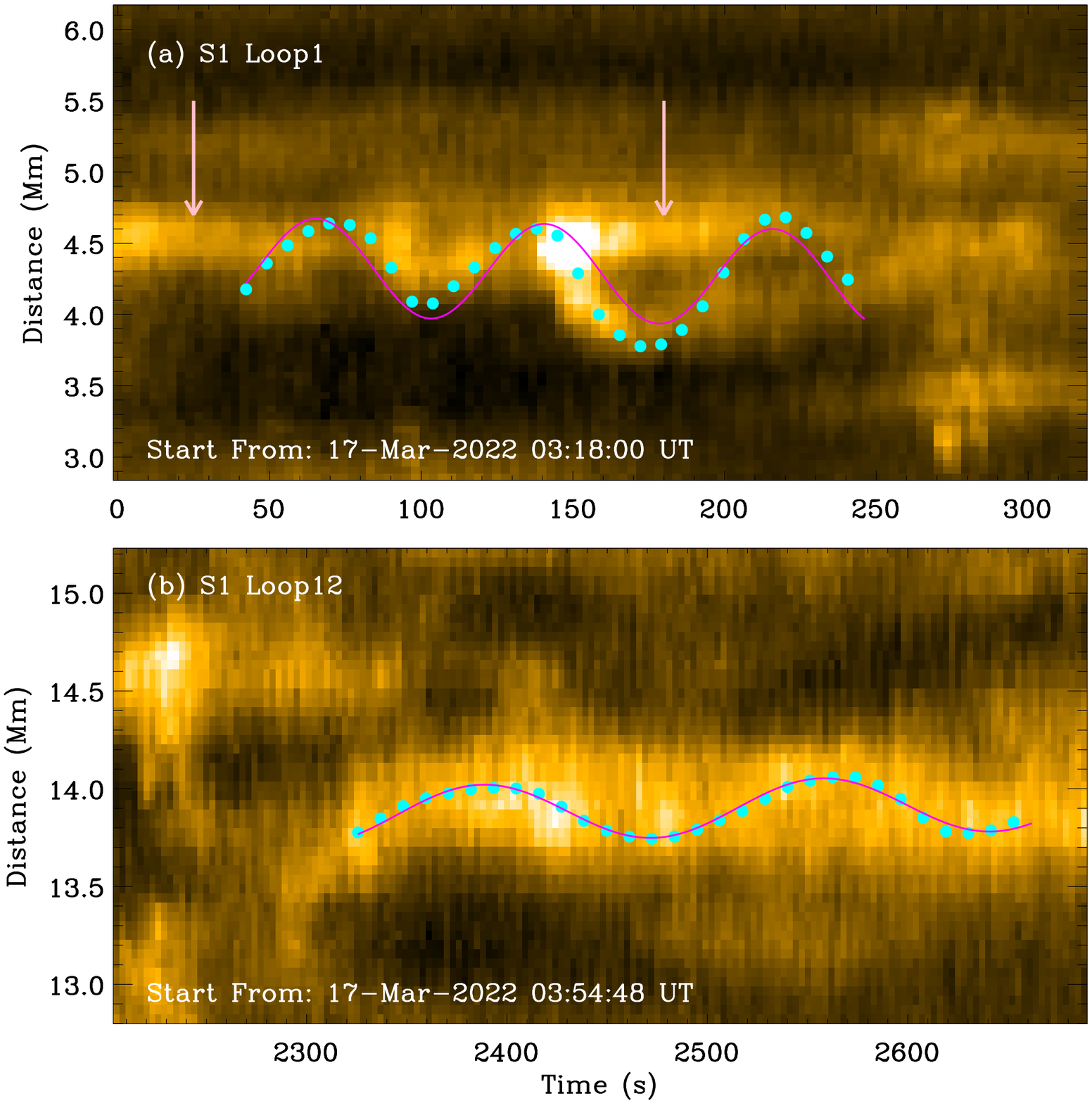}
\caption{Zoomed time-distance maps made from slit~S1 in region~R1
along the oscillating loops~1 (a) and 12 (b). The cyan circles
indicate the fitting profile positions of coronal loop oscillations,
and the magenta curves represent their best fitting results.
\label{fc_s1}}
\end{figure}

\begin{figure}
\centering
\includegraphics[width=0.8\linewidth,clip=]{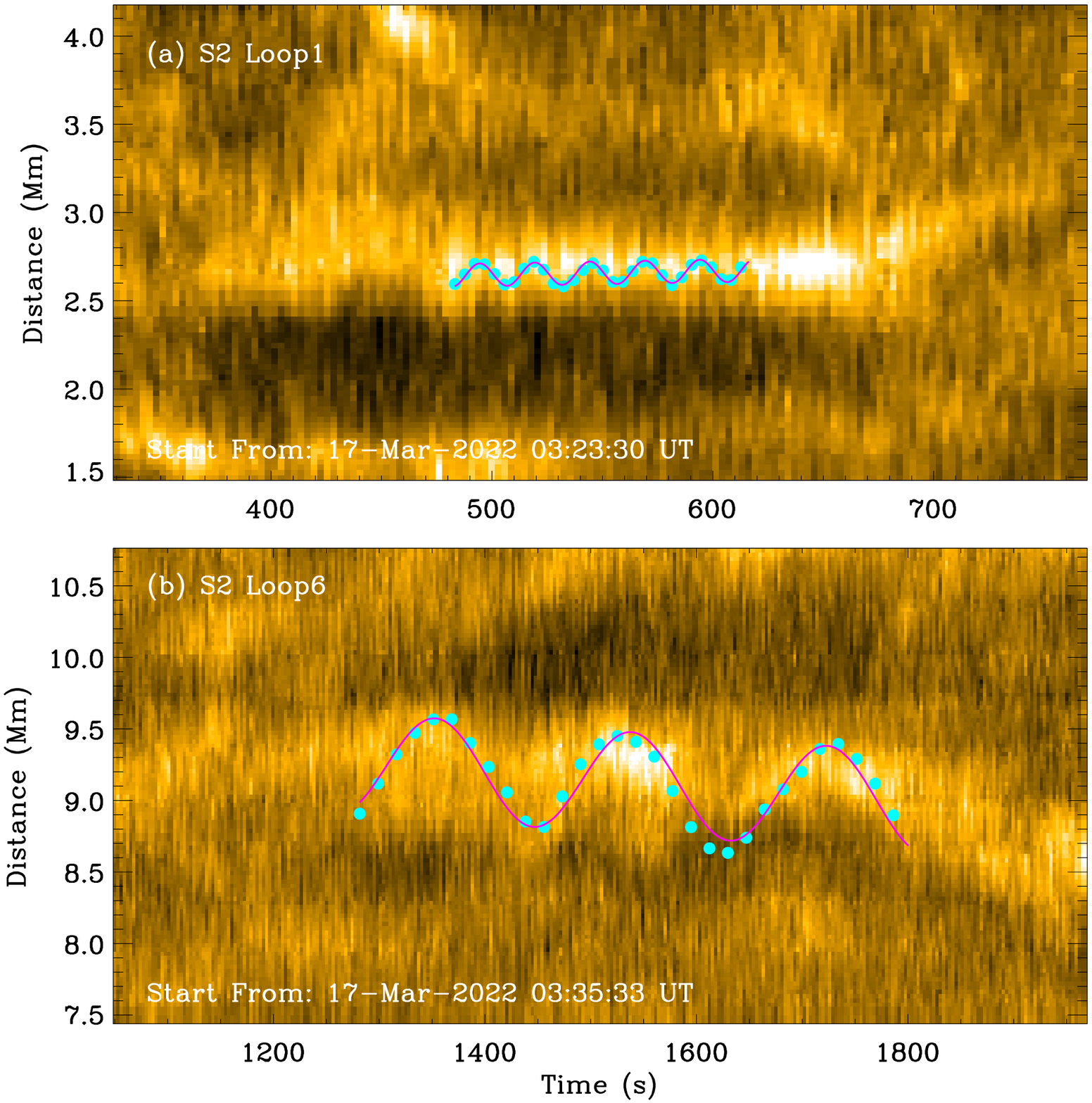}
\caption{Zoomed time-distance diagrams made from slit~S2 in
region~R1 along the oscillating loops~1 (a) and 6 (b). The cyan
circles indicate the fitting profile positions of coronal loop
oscillations, and the magenta curves represent their best fitting
results. \label{fc_s2}}
\end{figure}

\begin{figure}
\centering
\includegraphics[width=0.8\linewidth,clip=]{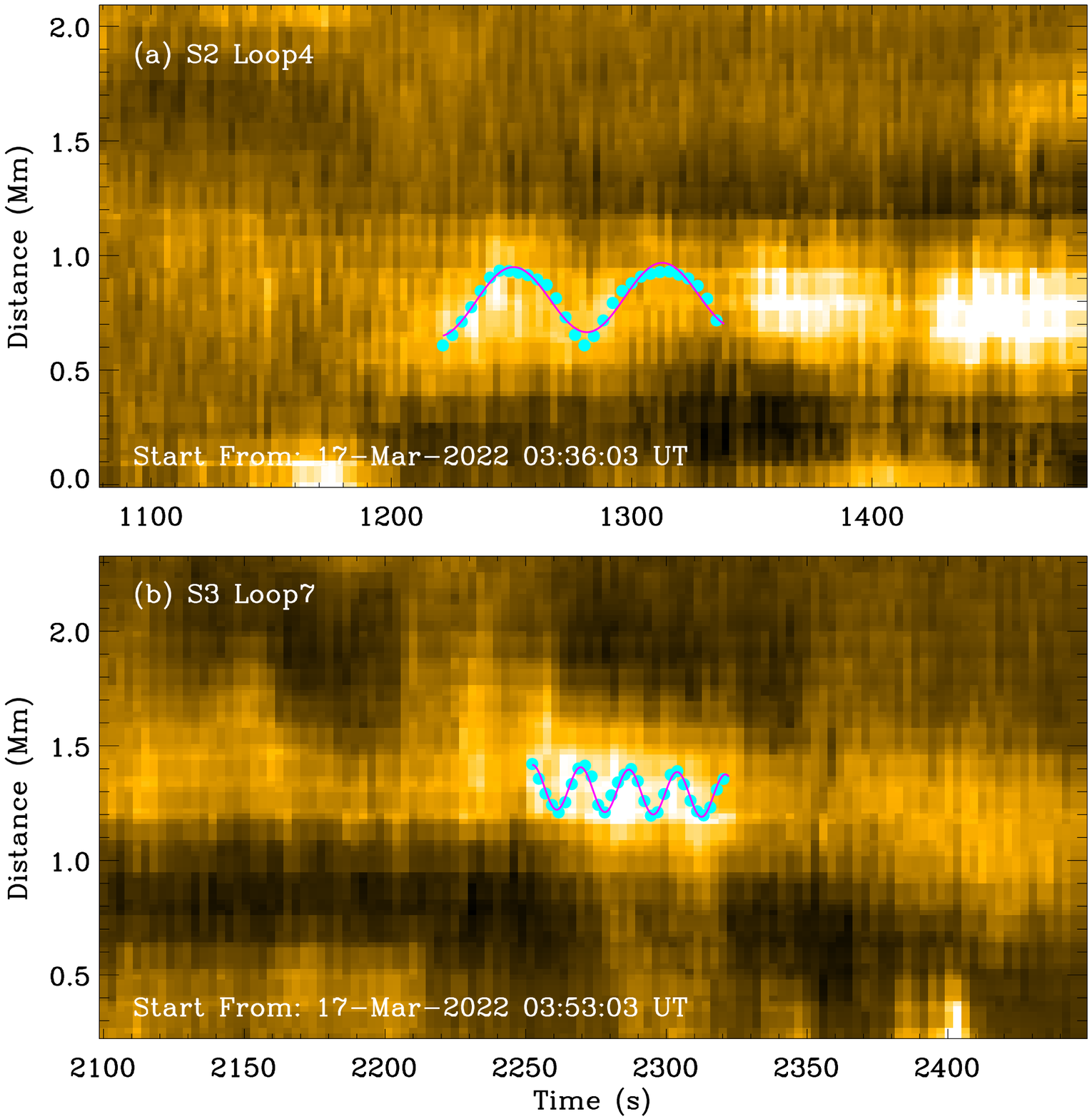}
\caption{Zoomed time-distance diagrams made from slit~S3 in
region~R2 along the oscillating loops~4 (a) and 7 (b). The cyan
circles indicate the fitting profile positions of coronal loop
oscillations, and the magenta curves represent their best fitting
results. \label{fc_s3}}
\end{figure}

\clearpage
\section{Energy estimations}
For the large-amplitude decaying oscillation, it was found that the
average decaying time is $\tau=1.79P$, where $P$ is the oscillation
period \citep[Section~3.7 in][]{Nechaeva19}. Based on the damping
mechanism, one can construct an equation as follows
\citep[cf.][]{Van14,Yuan16a}:
\begin{eqnarray}
  E(t) = \frac{1}{2}V(\rho_i~v_m^{2}+\frac{b^2}{\mu_0})e^{-2t/\tau},
\end{eqnarray}

\noindent where, $E(t)$ represents the total (kinetic+magnetic)
energy averaged over oscillation period, $V$ represents the volume
of oscillating loops, $\rho_i$ and $b$ are the plasma density and
the magnetic field perturbation inside oscillating loops, $v_m^2$
denotes to the velocity amplitude, $\mu_0$ is the magnetic
permittivity of free space, and $\tau$ is the decaying time measured
for the displacement amplitude.

Then, its derivative $\frac{dE(t)}{dt}$ at $t=0$ will give us an
estimate for the total oscillation energy losses ($E_t$), such as:
\begin{eqnarray}
  E_t = - \frac{V}{\tau}(\rho_i~v_m^{2}+\frac{b^2}{\mu_0}),
\end{eqnarray}

\noindent Here `-' means that the wave is losing energy. For
clarity, we ignore `-' in the next discussion and thus could regard
as its contribution to coronal heating.

Assuming that the coronal loop has a constant cross-section ($S$)
\citep[cf.][]{Williams21,Gao22}, and the loop length ($L$) could
also be measured, then we can get $V=SL$. Considering that
$c_k=2L/P$ and $\tau=1.79P \approx 2P$, the energy flux ($E$) can be
expressed as:
\begin{eqnarray}
  E = \frac{E_t}{S} \approx \frac{1}{4}c_k(\rho_i~v_m^{2}+\frac{b^2}{\mu_0}),
\end{eqnarray}

Next, assuming that the same damping mechanism with the same damping
rate operates in the decayless oscillation, but it is compensated
with continuous energy supply from an unknown driver, we can obtain
Equation~(5) in our work. Therefore, the energy flow in the case of
decayless oscillations looks like the following: ``unknown driver
$\rightarrow$ standing kink oscillations $\rightarrow$ rapid damping
$\rightarrow$ plasma heating''. Such process continues when an unknown
driver is existing, and the decayless oscillation could be detected
in the coronal loop.

\end{document}